\documentclass[preprint, aip, pof,showpacs]{revtex4-1} 
\usepackage{graphicx, bm, bbm,amsmath, amssymb, epsfig}

\pdfoutput=1
\newcommand{\Pint}{~P\hspace{-.40cm}\int}
\newcommand{\nn}{\noindent}
\newcommand{\no}{\nonumber}
\newcommand{\bq}{\begin{align}}
\newcommand{\eq}{\end{align}}
\newcommand{\etal}{{\it{et al.}}~}

\begin{document}
\title{Flag flutter in inviscid channel flow}
\author{Silas Alben}
\affiliation{$^1$University of Michigan}
\email{alben@umich.edu}

\date{\today}

\begin{abstract}
Using nonlinear vortex-sheet simulations, we determine the region in parameter
space in which a straight flag in a channel-bounded inviscid flow
is unstable to flapping motions. We find that for heavier flags,
greater confinement increases the size of the region of instability.
For lighter flags, confinement has little influence. We then
compute the stability boundaries analytically for an infinite flag,
and find similar results. For the finite flag we also consider
the effect of channel walls on the large-amplitude periodic flapping
dynamics. We find that multiple flapping states are possible but
rare at a given set of parameters, when periodic flapping occurs. As the channel
walls approach the flag, its flapping amplitude decreases roughly
in proportion to the near-wall distance, for both symmetric and
asymmetric channels. Meanwhile, its dominant flapping frequency
and mean number of deflection extrema (or ``wavenumber'') 
increase in a nearly stepwise fashion. That is, they remain nearly 
unchanged over a wide range of channel spacing, but when the
channel spacing is decreased below a certain value, 
they undergo sharp increases corresponding
to a higher flapping mode.
\end{abstract}


\maketitle
\section{Introduction}

A variety of experimental and theoretical studies have been conducted
on the flutter of flexible plates (or ``flags'') in recent years
\cite{kornecki1976ait,Dowell1980,Huang_JFluidsStruct_1995,ZCLS2000,
FP2001,watanabe2002esp,ZP2002,
TYD_JFluidsStruct_2003,SVZ2005,AM2005,
ESS2007,connell2007fdf,eloy2008aic,alben2008ffi,Michelin2008,manela2009}, 
following earlier work in the field of aeroelasticity 
\cite{Theodorsen1935,Fung1955,bisplinghoff2002pa}. Some of
the more recent studies, 
including extensions to multiple-flag
interactions and three-dimensional effects,
were reviewed by Shelley and Zhang \cite{shelley2011flapping}. 
Many of these studies addressed the stability problem: 
determining the region in parameter space where a flag in a 
uniform flow becomes unstable to transverse oscillations.
Many of the studies also characterized the flag
dynamics which occur after the instability grows to 
large-amplitude flapping, including the transitions from
periodic to chaotic motions \cite{AS2008,chen2014bifurcation}. 
In most cases flag flutter was studied in flows which
are unconfined or approximately so---e.g. in a flow
tunnel where the tunnel walls are far from the flag. 
Doare \etal considered the modification to flapping
due to spanwise confinement \cite{Doare2011,Doare2011a}. Guo 
and Paidoussis studied the flutter boundary for a
flag confined in the direction transverse to its
resting planar state \cite{Guo2000}, which is also the focus of
the present work. They studied plates with various
combinations of clamped, pinned, and free boundary conditions
at the leading and trailing edges. 
In the present work we focus on clamped and free
boundary conditions at the leading and trailing
edges, respectively, which were the most common
boundary conditions employed in 
several recent studies \cite{shelley2011flapping}.
We expand on the work of Guo 
and Paidoussis in a few ways.
First, they considered only the case of the flag support
point
placed symmetrically in the channel, while
here we consider both symmmetric and asymmetric
placement. Second, their focus was on the flutter
instability boundaries. Here we compute
these boundaries and also determine the large-amplitude
flapping behaviors, in the presence of vortex wake dynamics. 
Third, we develop an infinite-flag
model for which the stability boundary can be computed
analytically, extending similar work from the
unconfined case\cite{SVZ2005,alben2008ffi} to the
confined case.
We also note
the work of Jaiman \etal \cite{Jaiman2014} which briefly considered
the effects of channel walls together with fluid compressibility and
viscous skin friction on the flag stability boundary.

The organization of the paper is as follows. Section
\ref{sec:Model} presents the model for the nonlinear dynamics
of a flag in inviscid channel flow.  Section \ref{sec:FinFlag} 
presents results for the stability boundaries
and large-amplitude dynamics. Section \ref{sec:InfFlag}
specializes the model to the case of an infinite flag and
presents analytical stability boundaries for various channel wall
spacings. Section \ref{sec:Conc} presents
the conclusions.

\section{Model \label{sec:Model}}

\begin{figure}
  \centerline{\includegraphics[width=18cm]
  {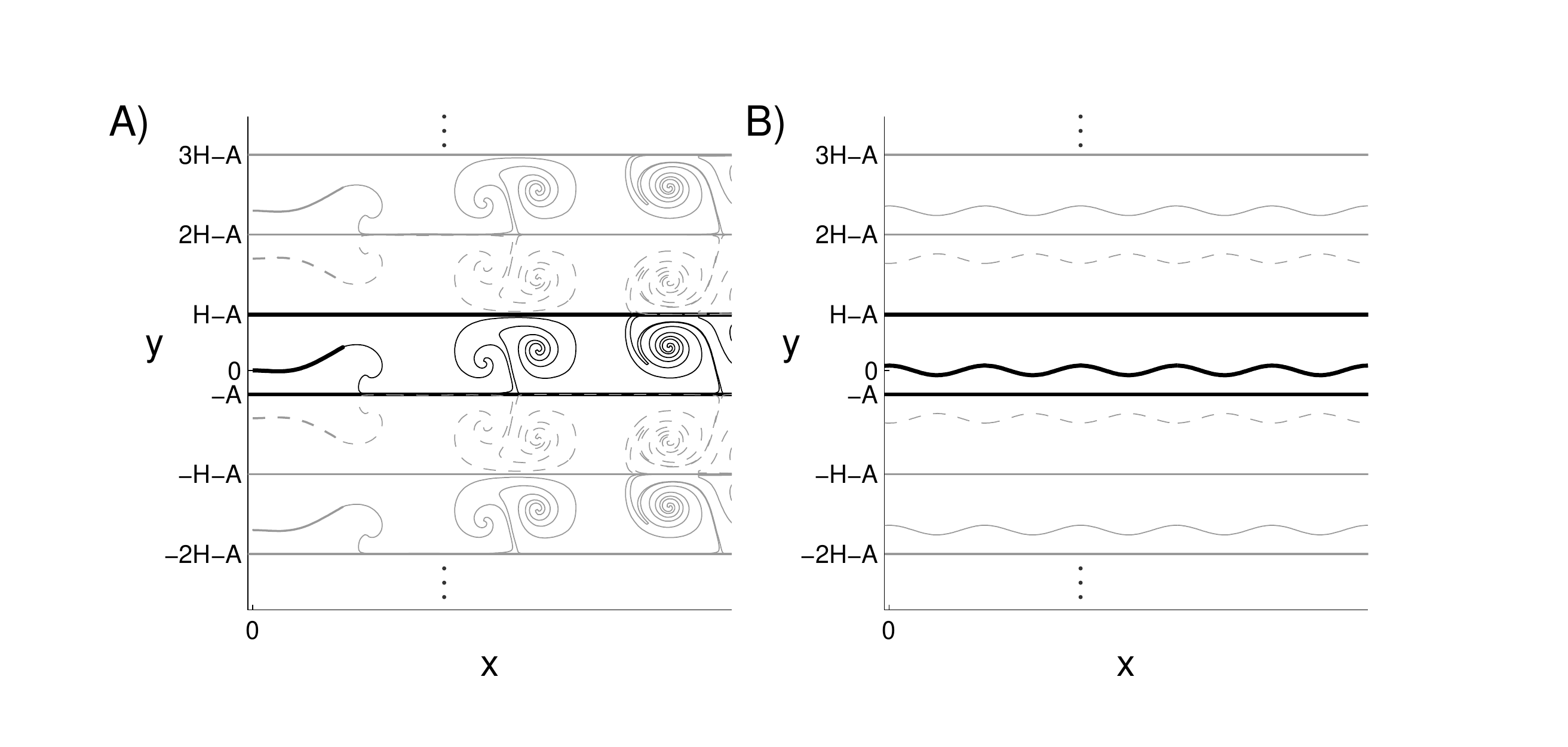}}
  \caption{Examples of image systems of flags and vortex wakes that 
impose no-penetration boundary conditions on the channel walls 
at $y = -A$ and $H-A$. A) A flag (thick black line emanating from
the origin) and
vortex wake (thin black line emanating from the flag's trailing edge) 
are shown with their images (solid and
dashed gray lines). B) For the infinite flag case, a sinusoidal body is
shown (thick black line) with images (solid and dashed gray lines).}
\label{fig:SchematicFlagChannel}
\end{figure}

We consider an inviscid model for a thin plate or flag oscillating
in a flow. The model is similar to some which were described previously 
\cite{AlbenJCP2009,AS2008}, so we will present it briefly, emphasizing the
modifications needed to incorporate the channel walls. 
An example of a flag and vortex wake in a channel
are shown in Fig. \ref{fig:SchematicFlagChannel}a. The channel is
the region $-A \leq y \leq H-A$. The flag is shown as a thick black line with
leading edge at the origin. The vortex wake is the thinner black line
that emanates from the trailing
edge of the flag and forms spirals as it evolves downstream of the flag.
A uniform horizontal flow with velocity 
$U\mathbf{e}_x$ has been applied at infinity upstream, 
and the flag, wake, and flow evolve under
a set of equations which we first summarize: 
Euler's equations of fluid momentum balance;
the no-penetration condition on the flag; a mechanical
force balance between flag bending rigidity, inertia, and fluid
pressure; Kelvin's Circulation theorem; the Birkhoff-Rott equation
for free vortex sheet dynamics; and the Kutta condition
governing vorticity production at the flag's trailing edge. We will
present the most important equations here and refer
to previous work\cite{AlbenJCP2009,Saffman1992} for the remainder and additional 
background information.
In the following
we nondimensionalize lengths by the flag length $L$, velocities by
the imposed flow speed $U$, and densities by the fluid density $\rho_f$.

The position of the flag is described as 
$\zeta(s,t) = x(s,t) + iy(s,t), 0 \leq s \leq 1$,
a curvilinear segment of length $1$ in the complex plane, parametrized by
arc length $s$ and time $t$. At each instant, the flow 
may be computed in terms of the position
and strength of a single vortex sheet in the plane. The vortex sheet has two parts:
a ``bound'' part, coincident with the flag itself, for $0\leq s \leq 1$, and
a ``free'' part, for $s > 1$, which emanates from the flag's trailing edge at $s = 1$.
On both parts, the vortex sheet's strength is denoted $\gamma(s,t)$ 
and its position is denoted $\zeta(s,t)$ 
(the same as the flag for $0 \leq s \leq 1$). On the bound
sheet, the vortex sheet evolves to satisfy the no-penetration condition
on the flag (but not the no-slip condition, since the flow is inviscid):
\begin{align}
\mbox{Im} \left(e^{-i\theta(s,t)}\, \partial_t \zeta(s,t)\right) &=
\mbox{Im} \left(e^{-i\theta(s,t)}\, \left(1+
\Pint_{0}^{L_w}~\gamma(s^{\prime},t) \overline{K(s,s^\prime,t)} \, ds^{\prime}
 \right)\right), \; 0 \leq s \leq 1.
\label{KinematicEqn}
\end{align}
\nn This condition sets the component of the body's velocity normal to the body equal
to the same component of the flow velocity. 
Here $\theta(s,t)$ is the tangent angle to the body. The unity term on the right
hand side is the uniform background flow and
$K(s,s^\prime,t)$ is the complex flow velocity at $\zeta(s,t)$ due
to a periodic array of point vortices of strength unity located at 
$\zeta(s^\prime, t) + 2inH, n = 0, \pm 1, \pm 2, \ldots$ as well as a periodic
array of point vortices of strength negative unity located at
$\overline{\zeta(s^\prime, t)} - 2iA + 2inH, n = 0, \pm 1, \pm 2, \ldots$. Together
the two
arrays give the flow in the channel due to a point vortex at $\zeta(s^\prime, t)$, 
enforcing the no-penetration conditions on the channel walls by the ``method of
images'' \cite{Saffman1992}. The special integral symbol in (\ref{KinematicEqn}) denotes a 
principal-value integral, due to 
the $\sim 1/(s-s^\prime)$ singularity in $K(s, s^\prime, t)$. The kernel
is given by
\begin{align}
K(s,s^\prime,t) = &\frac{1}{4H}\coth\left(
\frac{\pi(\zeta(s,t)-\zeta(s^\prime,t))}{2H}\right) \no \\
&-\frac{1}{4H}\coth\left(
\frac{\pi\left(\zeta(s,t)-\overline{\zeta(s^\prime,t)}+2iA\right)}{2H}\right), 
\; 0 \leq s,s^\prime \leq 1. \label{K}
\end{align}
\nn We use a regularized version of the kernel on the 
free vortex sheet ($s' > 1$), which allows for smooth vortex sheet dynamics,
analogous to Krasny's method \cite{Krasny1986}:
\begin{align}
&K(s,s^\prime,t) = \frac{1}{8H}\frac{\displaystyle \sinh\left(\frac{\pi}{H} (x(s,t) - x(s', t))\right)
-i \sin\left(\frac{\pi}{H} (y(s,t) - y(s', t))\right)}{
\displaystyle \sinh^2\left(\frac{\pi}{H} (x(s,t) - x(s', t))\right)
+ \sin^2\left(\frac{\pi}{H} (y(s,t) - y(s', t))\right) + 
\left(\frac{\pi \delta(s',t)}{2H}\right)^2} \no \\
&-\frac{1}{8H}\frac{\displaystyle \sinh\left(\frac{\pi}{H} (x(s,t) - x(s', t))\right)
-i \sin\left(\frac{\pi}{H} (y(s,t) + y(s', t)+2A)\right)}{
\displaystyle \sinh^2\left(\frac{\pi}{H} (x(s,t) - x(s', t))\right)
+ \sin^2\left(\frac{\pi}{H} (y(s,t) + y(s', t)+2A)\right) + 
\left(\frac{\pi \delta(s',t)}{2H}\right)^2}
, \no \\
& 0 \leq s \leq 1, \; 1 < s^\prime. \label{Kdelta}
\end{align}
\nn We set
\begin{align}
\delta(s',t) = \delta_0 \left(1-e^{\displaystyle -|\zeta(1,t)-\zeta(s^\prime,t))|^2/4\delta_0^2}\right)
\end{align}
\nn with $\delta_0$ = 0.2. This regularization tapers to zero at the trailing edge $s' = 1$,
so $K(s,s',t)$ is continuous there. In other words, as $\delta(s',t) \to 0$, the expression in
(\ref{Kdelta}) tends to that in (\ref{K}). The tapered regularization
allows for smooth vortex sheet dynamics away from the trailing edge while
decreasing the effect of regularization on the generation of
vorticity at the trailing edge \cite{alben2010regularizing}. At the
trailing edge, the vortex sheet is advected away from the body by the uniform background flow,
so it remains in the less regularized region near $s' = 1$ for a time which is too short to allow
chaotic dynamics to develop.

The vortex sheet strength $\gamma(s,t)$ is coupled to the pressure jump $[p](s,t)$ across the
flag using a version of the unsteady 
Bernoulli equation\cite{HLS_JComputPhys_2001,alben2012attraction}:
\begin{align}
\partial_t \gamma(s,t) + \partial_s\left((\mu(s,t) -\tau(s,t))\gamma(s,t)\right) = \partial_s [p](s,t), \;
0 \leq s \leq 1. \label{Bernoulli}
\end{align}
\nn Here $\mu(s,t)$ is the component of the flow velocity tangent to the body,
\begin{align}
\mu(s,t) =  \mbox{Re} \left(e^{-i\theta(s,t)}\, \left(1+
\Pint_{0}^{L_w}~\gamma(s^{\prime},t) \overline{K(s,s^\prime,t)} \, ds^{\prime}
 \right)\right)
\end{align}
\nn and $\tau(s,t)$ is the body's velocity component tangent to itself:
\begin{align}
\tau(s,t) =  \mbox{Re} \left(e^{-i\theta(s,t)} \partial_t \zeta(s,t)\right).
\end{align}

The unsteady Euler-Bernoulli beam equation couples the pressure loading
to body inertia and bending rigidity (for a uniform beam here):
\begin{align}
R_1 \partial_{tt} \zeta + R_2 \partial_s \left(\partial_s \kappa(s,t) ie^{i\theta(s,t)}\right)
-\partial_s\left(T(s,t)e^{i\theta(s,t)}\right) = -[p]ie^{i\theta(s,t)}. \label{beam}
\end{align}
\nn Here $R_1$ and $R_2$ are the dimensionless material parameters for the flag:
\begin{align}
R_1 = \frac{\rho_s h}{\rho_f L} \;, \; R_2 = \frac{B}{\rho_f U^2 L^3 W}
\end{align}
\nn Here $\rho_s$ is the mass per unit volume of the flag and $h$ is its thickness.
We assume that $h/L$ is small, but $\rho_s/\rho_f$ may be large, 
so $R_1$ may assume any nonnegative value. As stated previously, $\rho_f$ is the
mass per unit volume of the fluid and $L$ is the flag length. $B$ is the flag bending
rigidity, $U$ is the uniform background flow speed, and $W$ is the
out-of-plane width of the beam. The flow is assumed to be a 2D flow, so it is
uniform in the out-of-plane direction. In (\ref{beam}), $\kappa = \partial_s \theta$
is the beam's curvature and $T(s,t)$ is the tension in the beam, arising
from its inextensibility. $T$ is eliminated in favor of $\kappa$ by
integrating the tangential component of (\ref{beam}) from the free end of the beam
($s = 1$) where $T = 0$ (and $\kappa = \partial_s\kappa = 0$). 
The normal component of (\ref{beam}) is then used
to relate $[p]$ to the beam shape and motion given by $\zeta$ and $\kappa$, with
``clamp'' boundary conditions described below.
Further details are given in a previous work\cite{AlbenJCP2009}. We also refer the reader 
to this work for information on how the vorticity in the free vortex sheet is 
generated at the trailing edge using the Kutta condition, and advected downstream 
using the Birkhoff-Rott equation. In the Results sections, we use the total 
circulation in the free vortex sheet as an indicator of the type of dynamics (periodic,
chaotic, etc.). The total circulation is
\begin{align}
\Gamma(t) = \int_1^{L_w} \gamma(s,t)\, ds.
\end{align}

We evolve the flag and flow using the equations just presented, as an 
initial-boundary-value-problem, with the flag starting in the
horizontal state with uniform flow velocity $U \mathbf{e}_x$ at $t = 0$. 
The upstream edge of the flag ($s = 0$) is
held fixed at the origin ($\zeta(0,t) = 0$), and its tangent angle is
perturbed from horizontal sufficiently smoothly 
that accelerations are continuous at $t = 0$:
\begin{align}
\theta(0,t) = \theta_0 t^3 e^{-t^3}\;,\; t \geq 0. \label{pert}
\end{align}
\nn In the simulations we use two different perturbation magnitudes,
$\theta_0 = 0.005$ and 0.1. We then compute the flag and flow for $t \geq 0$.
For some parameters, as $t$ exceeds 1 the flag deflection 
decays exponentially with time, and
tends to a flat, horizontal state, which is
then considered to be a stable equilibrium. For other parameters, the flag deflection
grows exponentially with time, in which case the flat state is considered to be unstable. 
The perturbation (\ref{pert}) decays rapidly for $t > 1$, so it provides
an initial perturbation and does not apply
significant forcing to the flag at later times ($t \gg 1$) which are our focus here.

In the following
sections, we determine where in the parameter space the horizontal flag is unstable. When
it is unstable, we characterize the large-amplitude dynamics of the flag (amplitude,
dominant frequency, and typical wavenumber), and how they depend on the parameters.

\section{Finite flag results \label{sec:FinFlag}}

We find
that for most values of $R_1$, $R_2$, $A/L$, and $A/H$, the
two perturbations ($\theta_0 = 0.005$ and 0.1) lead to the 
same flapping state once the 
perturbation reaches an order-one magnitude. 
This is particularly true when the flapping state is 
periodic with a single dominant frequency. For smaller values of
$R_2$, chaotic flapping occurs with a broadband frequency
spectrum, and it is difficult to precisely define the final
flapping state. For simplicity we focus on periodic flapping
states.

\begin{figure}
  \centerline{\includegraphics[width=13cm]
  {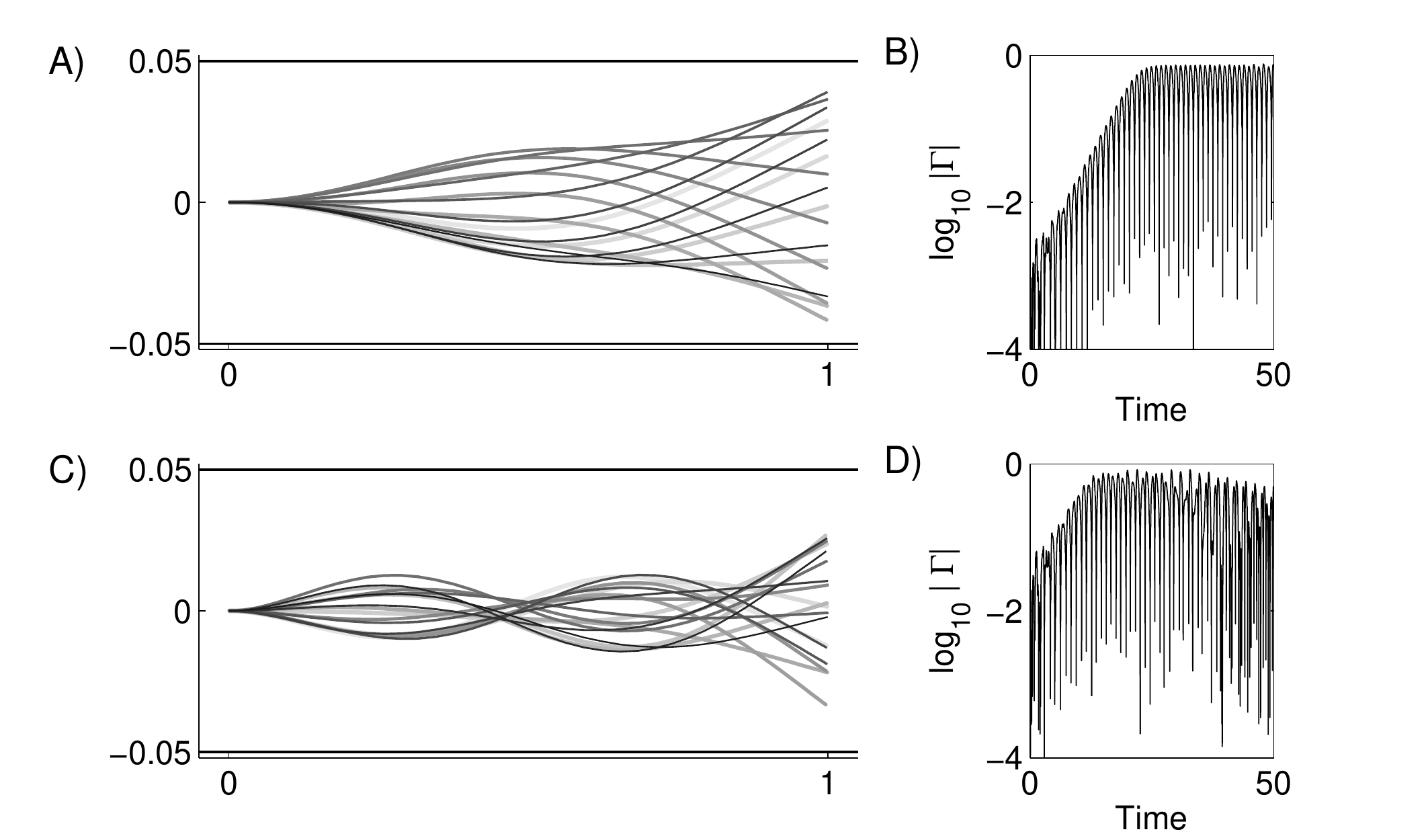}}
  \caption{Examples of different flapping states which are
obtained at the same parameter values but with
different initial perturbations.  
For a smaller
perturbation, $\theta_0(t) = 0.005\,t^3 e^{-t^3}$, flag snapshots (A)
and total wake circulation versus time (B) are shown. For a larger
perturbation $\theta_0(t) = 0.1\,t^3 e^{-t^3}$, the corresponding
results are shown in (C) and (D). For both cases, $R_1 = 2.11$, $R_2 = 0.0395$,
$A/L = 0.05$, and $A/H = 0.5$. For A and B, the channel walls are plotted
with solid black lines. Note that the axes are not to scale.}
\label{fig:MultipleFlappingStates}
\end{figure}

In figure \ref{fig:MultipleFlappingStates}, we present an example
in which our two differently-sized initial perturbations lead
to apparently different flapping states, for a flag in a symmetric
channel ($A/H = 0.5$) of fairly small width ($A/L = 0.05$).
The eventual flapping state for the
smaller perturbation, shown in panel a, is periodic with a single
dominant frequency. The magnitude of the total wake 
circulation $|\Gamma|$, plotted in panel b, rises exponentially in
time before saturating as a periodic oscillation. The larger initial
perturbation leads to a higher wavenumber flapping mode, shown in panel c.
The flapping is also much less regular, and the dominant frequency is
less dominant. This can be seen qualitatively by comparing the flapping
dynamics (panel c) and
circulation dynamics (panel d) with the quasi-periodic case (panels a and b).

We now proceed to characterize the finite flag's dynamics across
parameter space.
As noted above, in most cases of periodic flapping, 
the same flapping state is obtained with
different initial perturbations. This is consistent with the observation
of a very small range of hysteresis and bistability
(about 1--2\% in $R_2$) in a previous study of
the unbounded flapping flag \cite{AS2008}. Therefore, for simplicity 
we proceed using the perturbation $\theta_0 = 0.1$ only, and focus on
the periodic states which arise with this perturbation, which are
almost always the same as those which 
arise with the perturbation $\theta_0 = 0.005$. 

\begin{figure}
  \centerline{\includegraphics[width=17cm]
  {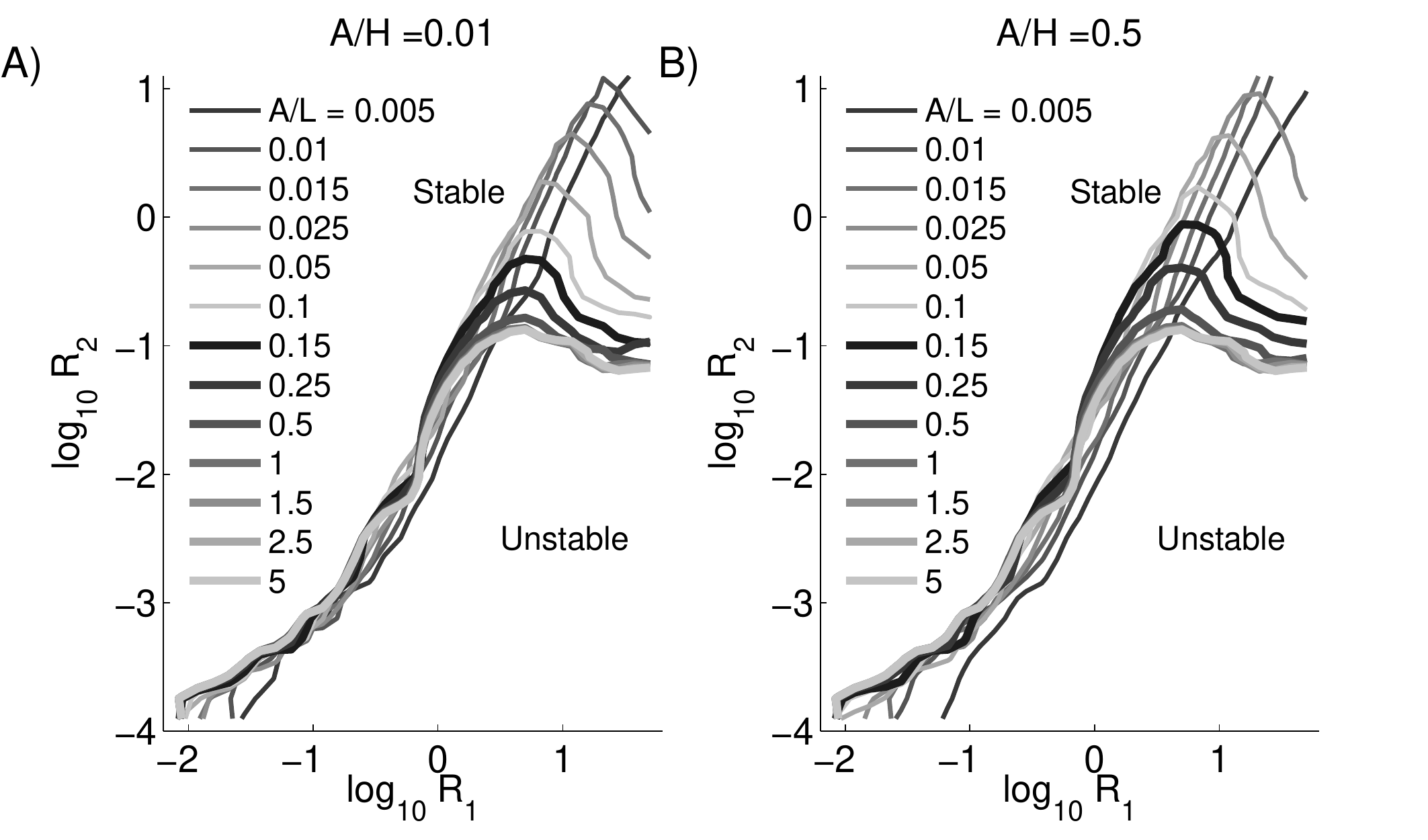}}
  \caption{Stability boundaries for finite flags in the space of
dimensionless flag inertia ($R_1$) and bending rigidity ($R_2$).
Boundaries are shown for various ratios of near-wall-distance
to flag length $A/L$ (labeled) when the far wall is much farther
away ($A/H = 0.01$, panel a) and when the two walls are equidistant 
($A/H = 0.5$, panel b).}
\label{fig:StabBdiesFig}
\end{figure}

We begin by computing the stability boundaries for the finite flag 
in the $R_1$-$R_2$ plane. We compute the growth (or decay) rates of the
initial perturbation from data analogous to those in Fig.
\ref{fig:MultipleFlappingStates}b, on a 33-by-33 grid of
values in $R_1$-$R_2$ space spanning several decades in each
parameter. We interpolate these values to find the location of
the line with zero growth rate, which is the stability boundary.
We compute the stability boundaries for two different values of $A/H$, 0.01 
(shown in Fig. \ref{fig:StabBdiesFig}a), for which the far wall
is 99 times as distant as the near wall, and 0.5
(shown in Fig. \ref{fig:StabBdiesFig}b), for a symmetric
channel. In each panel, the stability boundaries are given for
twelve different values of near-wall-distance ($A/L$) ranging from 
0.005 to 5 (labeled), and plotted with different shades of gray and line
thicknesses. 

For the larger $A/L$ (1.5, 2.5, and 5), we find good convergence to
the stability boundary for the unbounded case in panels a and b. 
As $A/L$ decreases, the
``plateau'' portion of the
stability boundary moves upwards, 
towards larger $R_1$ and $R_2$, and becomes more curved. 
At smaller $R_2$, the stability boundary
shifts slightly rightward, towards larger $R_1$, mainly at very
small values of $A/L$.
The main difference between panels a and b is that the shifts
are somewhat greater in panel b, consistent with an increased
instability due to increased confinement.

We now consider the large-amplitude flapping dynamics, 
and use three quantities to characterize them.
One is the frequency, defined as the maximum frequency in the power spectrum,
computed using Welch's method \cite{welch1967use}. We compute the
power spectrum from a plot of the circulation versus time, 
after the first time at which the circulation magnitude has 
reached 95\% of its global maximum up to $t = 50$ (in units of
$L/U$).  We present frequency (and other) data for cases 
in which the magnitude of the dominant
frequency peak is at least 10 times that of the second highest peak, so
that a dominant frequency can be identified clearly.
The second quantity is the amplitude, defined as the largest vertical deflection 
of the flag over the time over which the power spectrum is computed.
The third quantity is the time-averaged number of local extrema in
vertical position along the flag, computed with the same temporal
data as the preceding quantities. This third quantity gives a measure
of the ``waviness'' of the flag, and is proportional to the wavenumber for
a sinusoidal flag shape.

\begin{figure}
  \centerline{\includegraphics[width=18cm]
  {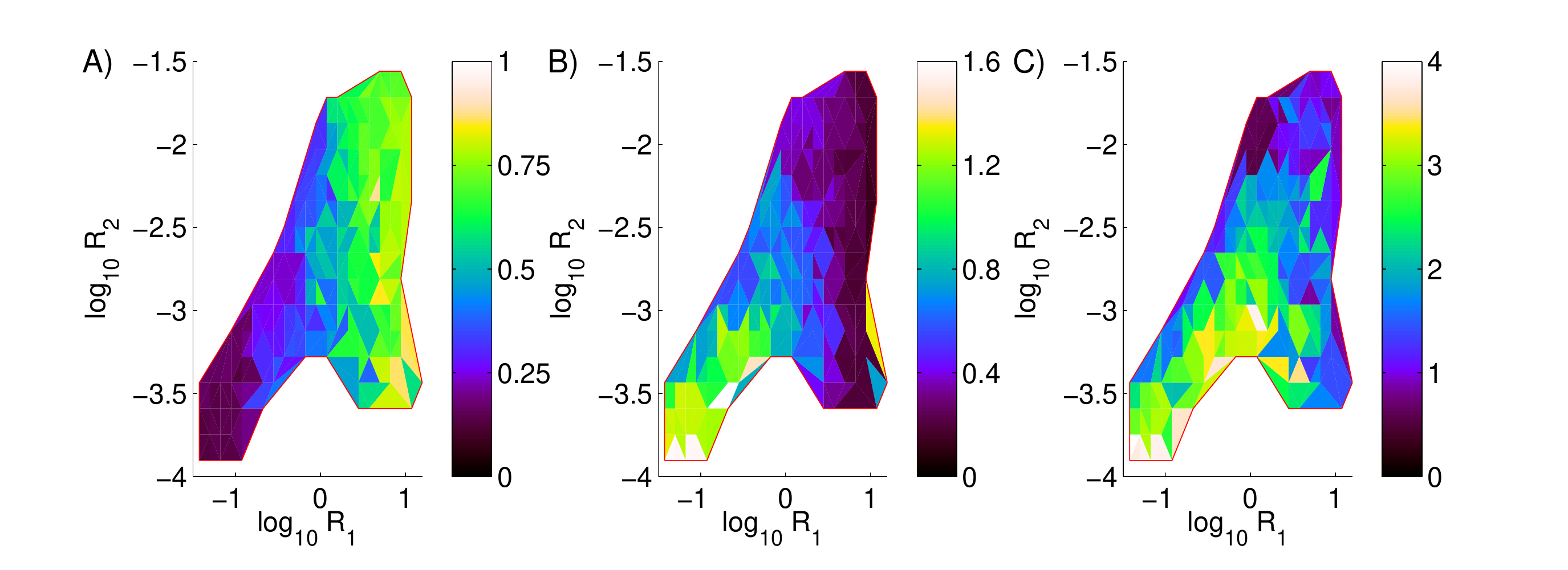}}
  \caption{Flapping amplitude (A), dominant frequency (B), and time-averaged
number of extrema in flag deflection (C), for an unbounded flag, in
the space of
dimensionless flag inertia ($R_1$) and bending rigidity ($R_2$). }
\label{fig:FlappingUnBdd}
\end{figure}

Before considering the effects of channel boundedness, we plot
in Fig. \ref{fig:FlappingUnBdd} the flapping amplitude (panel a),
frequency (b), and time-averaged number of extrema (c) for the unbounded
case, in the $R_1$-$R_2$ plane. This provides a baseline for
comparison with the channel-bounded cases. We focus on the region
where reliable frequency data can be obtained. This is the region
outlined in red in panels a, b, and c. Above and to the left of
this region, the straight state is stable and no flapping occurs. 
To the right of this region, the flag inertia $R_1$ is so large
that simulations take a very long time to converge to a stable
quasi-periodic state. Below this region, the flapping is too 
chaotic to have a clear definition of a dominant frequency. 
Nonetheless, the red region covers
a large portion of the space of flapping states, and certain
basic trends can be seen.

In panel a, we see that the amplitude increases with increasing 
$R_1$. Larger flag inertia results in larger
flag momentum. The flag is able to maintain its momentum for longer
times against fluid forces, and moves farther before reversing
direction. The amplitude does not vary as strongly
with bending rigidity ($R_2$). 
In panel b, we see that the dominant frequency decreases with
increasing flag inertia. This is again because the flag 
maintains its momentum for longer times, increasing the flapping
period and decreasing the frequency. The frequency also
decreases with increasing bending rigidity. Bending rigidity
also resists rapid changes in flag position under
changing fluid forces. This trend is opposite to
that for a beam oscillating in a vacuum, so clearly
fluid forces play an essential role. 
In panel c, we see that the average number of
maximum deflections along the flag increases as the
flag becomes more flexible ($R_2$ decreases), as expected---a
more flexible flag adopts a ``wavier'' shape. At a given
$R_2$, the number of maxima first increases, then
decreases as $R_1$ increases. This behavior does not have an
explanation which is obvious to us.

\begin{figure}
  \centerline{\includegraphics[width=13cm]
  {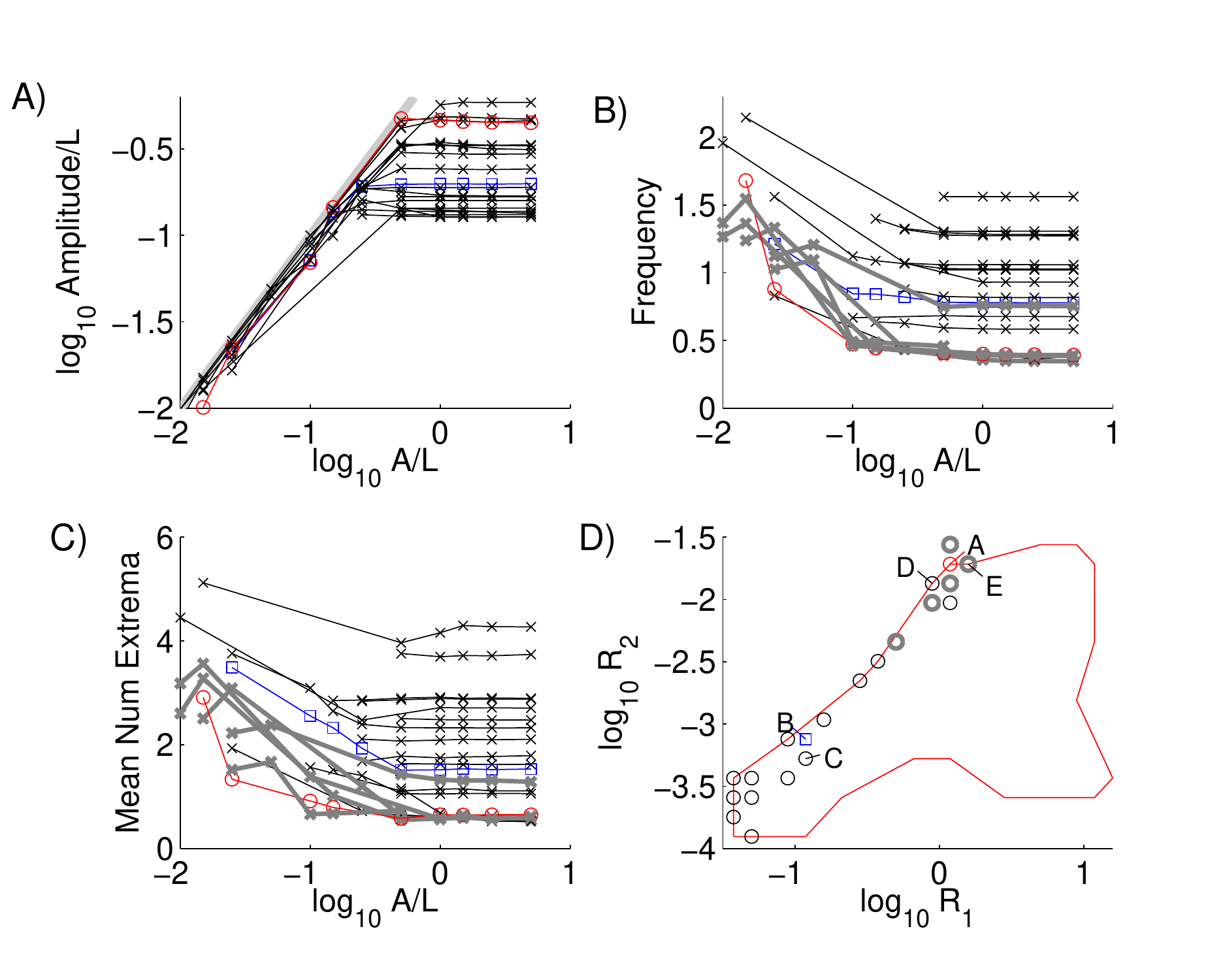}}
  \caption{Flapping amplitudes (A), dominant frequencies (B), and time-averaged
numbers of extrema in flag deflection (C) for flags in symmetric
channels ($A/H = 0.5$) with various wall spacings ($A/L$, shown on
horizontal axes). The crosses connected by a given line correspond to
a given $(R_1, R_2)$ pair. These are not labeled individually but the
ensemble of values is shown by circles (and other symbols) 
in panel D. For five values 
with labels A-E, flag snapshots are shown in the
panels of figure \ref{fig:FinalVaryAFixedAH} with corresponding
labels.}
\label{fig:VaryAFixedAH}
\end{figure}

\begin{figure}
  \centerline{\includegraphics[width=16cm]
  {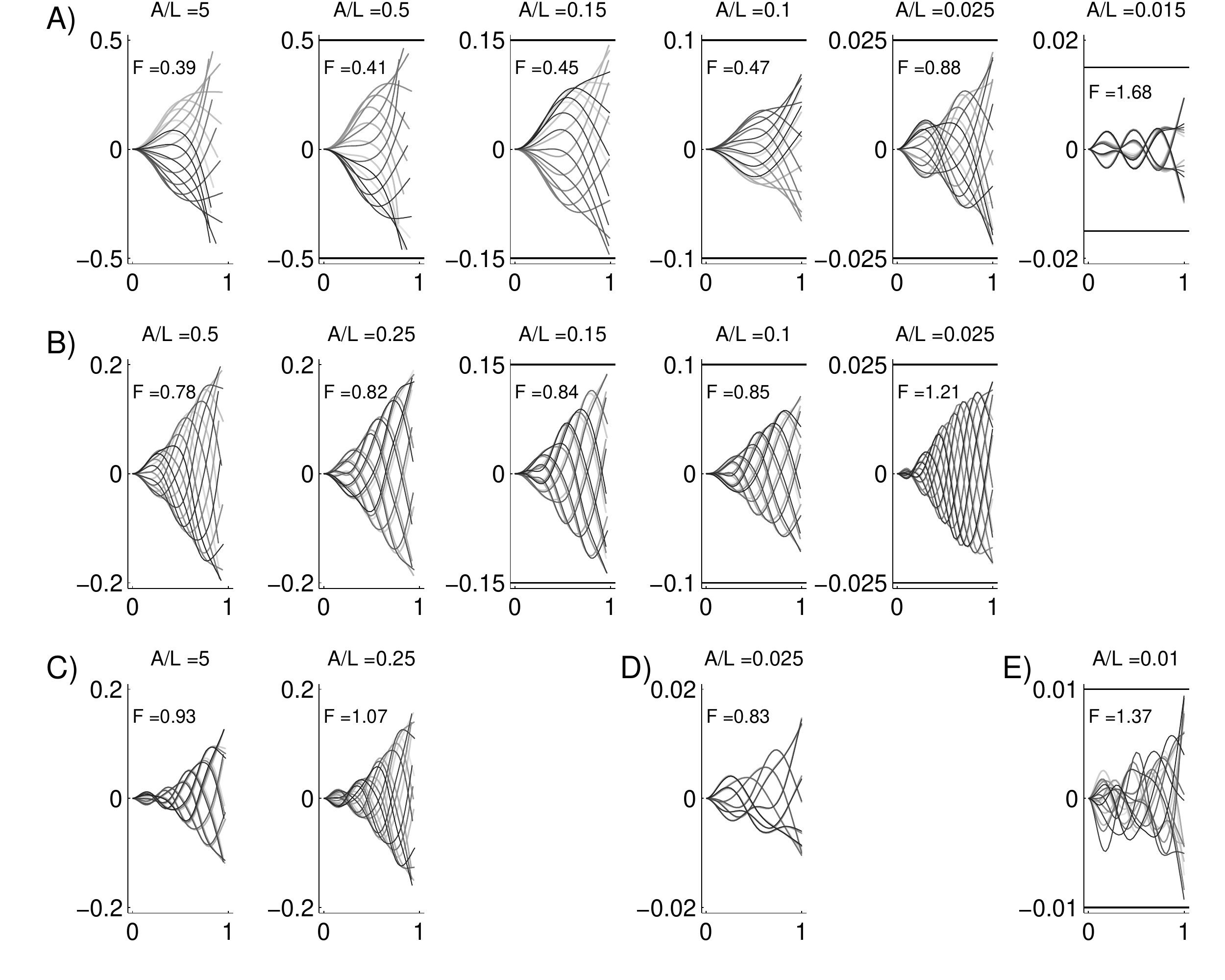}}
  \caption{Snapshots for flags with fixed values of $R_1$ and
$R_2$ in symmetric channels ($A/H = 0.5$) of varying half-width ($A/L$). 
Each of the panels A-E corresponds to a different pair of $R_1$ and $R_2$, labeled 
in figure \ref{fig:VaryAFixedAH}D. Within each panel, a frame shows flag snapshots
at a particular $A/L$ (labeled). The axes have different scales.
The dominant flapping frequencies are labeled F.}
\label{fig:FinalVaryAFixedAH}
\end{figure}

Having described how some of the basic quantities of the
flapping state---amplitude,
frequency, and number of deflection extrema---vary with
$R_1$ and $R_2$ in the unbounded state, we now consider how these
quantities vary with the channel geometry parameters, $A/L$ and
$A/H$. First we consider the symmetric channel ($A/H = 0.5$). We
identify a set of $(R_1, R_2)$ which have a dominant flapping
frequency across a wide range of $A/L$. These values of 
$(R_1, R_2)$ are shown by circles 
in Fig. \ref{fig:FinalVaryAFixedAH}d. For each such
$(R_1, R_2)$, we plot the flapping amplitude (panel a),
frequency (panel b) and average number of extrema (panel c).
In each of the panels a--c, values for a given $(R_1, R_2)$
are a set of crosses connected by a solid line. We
do not label each line by its value of $(R_1, R_2)$ to
prevent visual clutter, and also because we are mainly
interested in the trends which are common across the
set of plots.

First, we consider the flapping amplitude (panel a). For most values of
$(R_1, R_2)$, the behavior is as follows. Starting at the right
side of the panel, where each line has its unbounded value
(described in figure \ref{fig:FlappingUnBdd}) at large $A/L$, we
move to the left, in the direction of decreasing $A/L$.  In most
cases the amplitudes are nearly unchanged (but with
a slight increase, evidence of attraction to the walls) 
until $A$ decreases
to about the size of the amplitude in the unbounded case. Then
the amplitude is forced to become at least as small as $A$.
In this regime, the amplitude is usually quite close to $A$, its
upper limit. But in some cases it is significantly less, so
the flag remains well away from the walls. Meanwhile,
the frequency (panel b) increases as the walls move towards
the flag. In several cases the increase occurs as a jump
separating approximate ``plateau'' regions. 
These cases are plotted with thick gray lines
in panel b, with $(R_1, R_2)$ shown as thick gray circles
in panel d. As the frequency increases,
so does the mean number of deflection extrema (panel c), showing
that the flag adopts a ``wavier'' shape as the walls move inward.
The thick gray lines correspond to those for the frequency
(panel b), showing that the mean numbers of extrema also
undergo jumps, and remain close to certain values over a range
of $A/L$. In panel d, five $(R_1, R_2)$ values are 
labeled with letters A-E.
These refer to the panels in Fig. \ref{fig:FinalVaryAFixedAH},
in which we show sequences of snapshots at these selected
$(R_1, R_2)$ values.

Fig. \ref{fig:FinalVaryAFixedAH}a shows a typical sequence
of flapping states as $A/L$ decreases for fixed flag and flow parameters
($R_1$ and $R_2$). In each panel the flapping frequency is listed, denoted
by F. First, at $A/L = 5$, the state is essentially the unbounded case.
In the next frame (moving to the right), $A/L = 0.5$, so the
walls are now quite close to the
flag. The flag motion is nearly
unchanged, however, with only a slight increase in amplitude. This
slight ``wall attraction'' occurs at various $(R_1, R_2)$. Moving
again to the right ($A/L = 0.15$), the flag continues
to flap with nearly the same shape, but with a much smaller amplitude,
enforced by the walls. It is almost as though the flapping motion has simply
been scaled down in the vertical direction. The flapping frequency is
also nearly the same. In the next frame ($A/L = 0.1$), the shape
and frequency are again nearly the same, but the amplitude has decreased
again. Interestingly, the flag is farther from the walls in a
relative sense than in the previous panels, even though the
degree of confinement is now greater. Moving to the next panel
($A/L = 0.025$), the flag finally undergoes a large change
in shape and in frequency (which is nearly double the previous value).
The flag has transitioned to a very different flapping state.
Another large transition occurs in the next frame ($A/L = 0.015$), 
where the frequency nearly doubles again and the number of extrema
also increases. Relative to the channel width, the flag is now farther from
the channel wall than in the previous panel. The data corresponding
to case A are shown by red circles in the panels of 
Fig. \ref{fig:VaryAFixedAH}.

Moving to panel b, the second row of Fig. \ref{fig:FinalVaryAFixedAH},
$R_1$ and $R_2$ (labeled `B' in Fig. \ref{fig:VaryAFixedAH}d) are now
reduced by about a factor of 10. In the unbounded state, similar
to the first frame, the flag shows a higher bending mode than in
panel a, and a higher frequency. As $A/L$ is decreased to 0.25, there
is essentially no change in the flapping state. As $A/L$ is
decreased again to 0.15 and to 0.1, the flapping amplitude drops
twice to fit within the successively smaller channels, but the
flag shapes and frequencies are only slightly changed. In the last
frame ($A/L = 0.025$), the flag has moved to a higher bending mode with
a higher frequency. The data corresponding
to case B are shown by blue squares in the panels of 
Fig. \ref{fig:VaryAFixedAH}.

Panel c shows, at a nearby value of $(R_1, R_2)$, another example of the
slight increase in flapping amplitude which can occur when the channel
walls move closer to the flag. The first frame ($A/L = 5$) is 
representative of the unbounded state. In the second ($A/L = 0.25$), the
walls are closer to the flag, and it has a small but noticeable increase 
in amplitude. Panel d shows an example of an asymmetric periodic state which occurs
even though the channel is symmetric. Such asymmetric states are common
in chaotic flapping, for which an example is shown in panel e, but are
rare for periodic flapping in the symmetric channel.

\begin{figure}
  \centerline{\includegraphics[width=13cm]
  {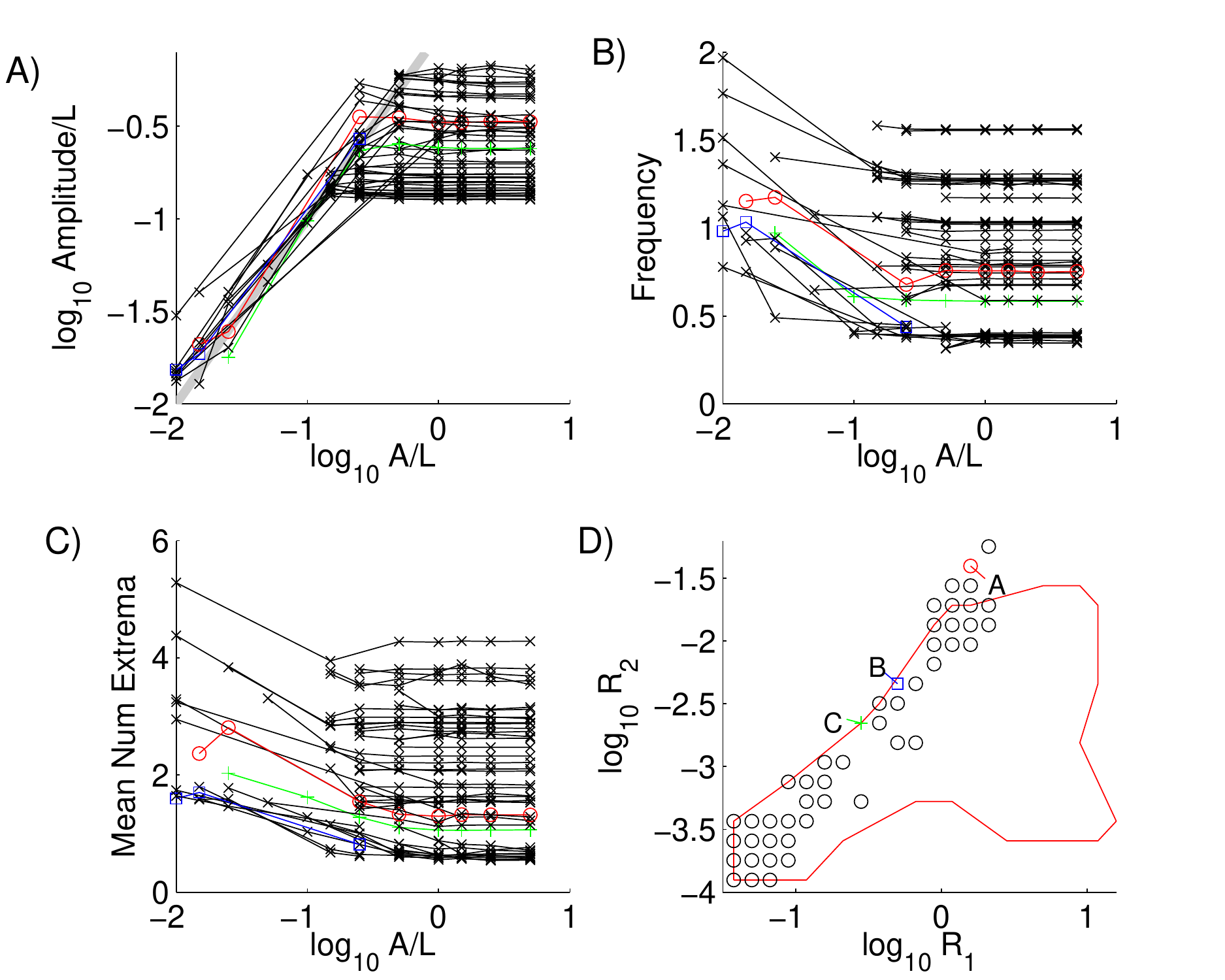}}
  \caption{Flapping amplitudes (A), dominant frequencies (B), and time-averaged
numbers of extrema in flag deflection (C) with various near-wall-spacings 
($A/L$, shown on horizontal axes), when the far wall is much
farther away ($0.01 \leq A/H \leq 0.1$). The crosses connected by a given line correspond to
a given $(R_1, R_2)$ pair. These are not labeled individually but the
ensemble of values is shown by circles (and other markers) 
in panel D. For three values 
with labels A-C, flag snapshots are shown in the
panels of figure \ref{fig:FinalVaryASmallAH} with corresponding
labels. Data for case A are shown with
red circles, for B with blue squares, and for C with
green plusses.}
\label{fig:VaryASmallAH}
\end{figure}

\begin{figure}
  \centerline{\includegraphics[width=16cm]
  {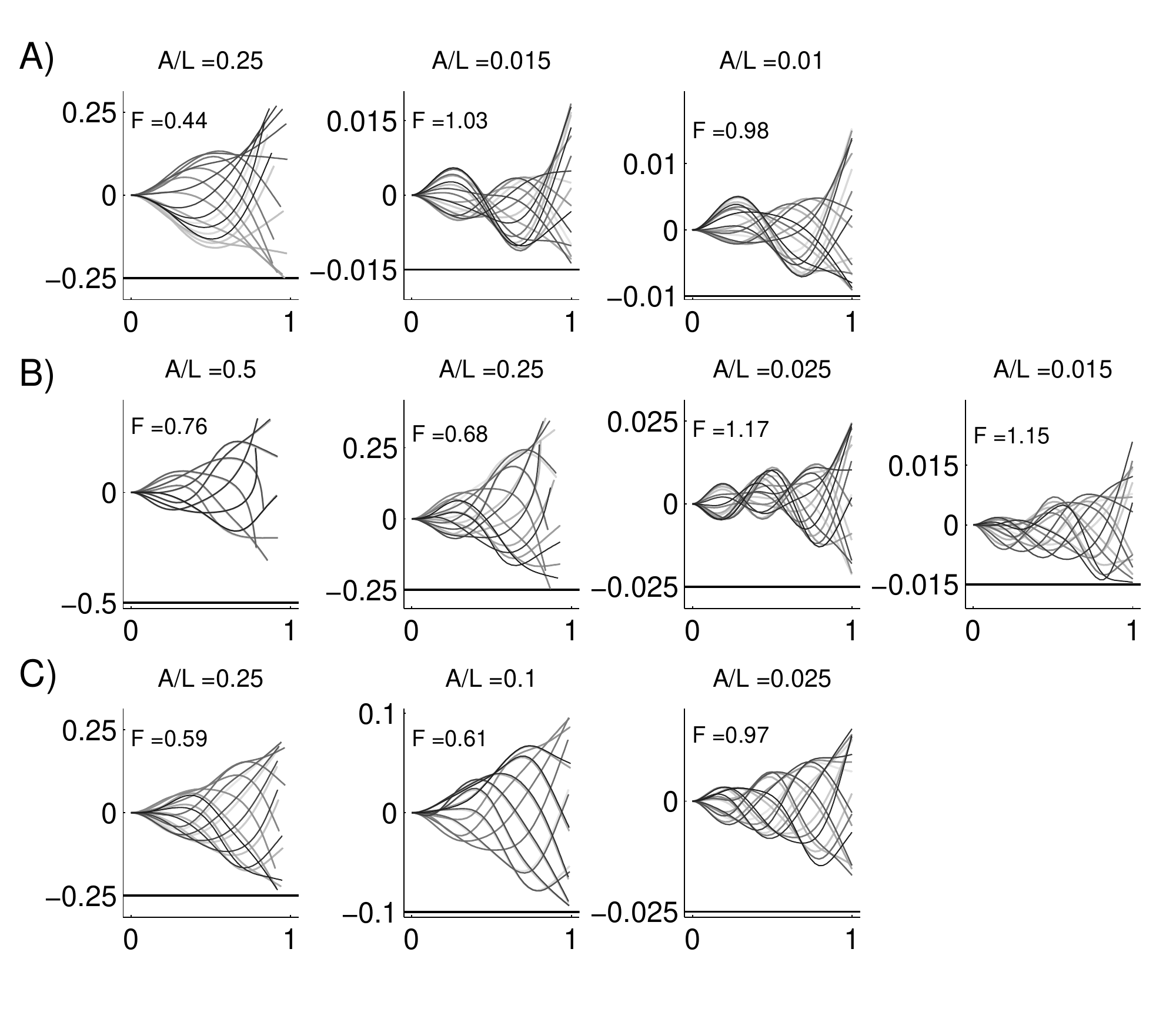}}
  \caption{Snapshots for flags with fixed values of $R_1$ and
$R_2$ with various near-wall-spacings 
($A/L$), when the far wall is much
farther away ($0.01 \leq A/H \leq 0.1$). 
Each of the panels a--c corresponds to a different pair of $R_1$ and $R_2$, labeled 
in figure \ref{fig:VaryASmallAH}d. Within each panel, a frame shows flag snapshots
at a particular $A/L$ (labeled). The axes have different scales.
The dominant flapping frequencies are labeled F.}
\label{fig:FinalVaryASmallAH}
\end{figure}

So far we have considered the large-amplitude dynamics of flags in symmetric
channels. We now look at the same results for flags when one channel wall
is much closer to the flag than the other. In Fig. \ref{fig:VaryASmallAH},
we show the flapping amplitudes (a), frequencies (b) and mean numbers of deflection
extrema (c) for an ensemble of flags with $0.01 \leq A/H \leq 0.1$, so the
near wall is a factor of 9--99 closer to the flag's leading edge
than the far wall. Once $A/H$ drops below 0.1, the results vary only
slightly (by three percent or less) with $A/H$. The trends are similar
to those in the symmetric case, with a few minor differences. In panel a,
as $A/L$ decreases (so the near wall gets nearer), 
the flapping amplitudes undergo a larger increase before they begin to
decrease. The reason is that the flags are mainly confined on one side,
so they may adopt an asymmetric flapping mode which flaps with larger
amplitude on the less constrained side. For this reason, the amplitude
now exceeds $A/L$ in most cases when $A/L$ becomes small. However,
the amplitude still scales approximately as $A/L$, meaning that the
degree of asymmetry in the vertical displacement remains bounded. Although
the flag could move to a much greater distance in the far-wall direction,
in most cases its maximum displacement in that direction is less than
50\% greater than its maximum displacement in the near-wall direction.
One might have expected the flag to flap closer to the near wall than
away from it
in some cases, in light of the slight wall attraction we observed
in the symmetric case. We have not observed such a case. If it occurs,
it probably occurs only slightly and only for a small range of $A/L$.
In panel d,
we again mark the $(R_1, R_2)$ values corresponding to the data 
presented. Three points are labeled A--C, and these refer to the
panels in Fig. \ref{fig:FinalVaryASmallAH} where the
corresponding snapshots are presented. In the panels
of Fig. \ref{fig:VaryASmallAH}, the data for case A are shown with
red circles, for B with blue squares, and for C with
green plusses. Cases A and B are examples in which
the frequencies and mean numbers of extrema show
patterns of jumping from one distinct mode to another, 
as in the symmetric cases.

Snapshots for case A are shown in 
Fig. \ref{fig:FinalVaryASmallAH}a. The flapping
is only slightly asymmetrical in the first frame ($A/L = 0.25$),
even though the flag nearly touches the near wall. In the second
frame ($A/L = 0.015$), the flag has moved to a clearly asymmetrical
mode, with the frequency nearly doubled. In the third frame 
($A/L = 0.01$), the mode and frequency are nearly the same,
but the amplitude is smaller. For case B (panel b), the first
two frames ($A/L = 0.5$ and 0.25) show a similar flapping mode with 
similar frequencies,
although the second is more irregular (or chaotic) in frequency
space. The third frame ($A/L = 0.025$) shows a switch to a different
mode and frequency. The flapping is however more periodic 
and less chaotic than in the second frame. In the fourth frame
($A/L = 0.015$), the mode and frequency are 
similar to those in the third, but now the flapping is more
chaotic. This alternating pattern of regular and chaotic flapping
has been seen at other $(R_1, R_2)$ as the walls become closer.
For case C (panel c), the first two frames ($A/L = 0.25$ and 0.1) show 
essentially the same
mode and frequency, though the amplitudes are very different.
Both motions are quite symmetrical despite the large channel
asymmetry. The third frame ($A/L = 0.025$) has a higher frequency
and number of extrema, and the degree of asymmetry is only
slightly increased.

\section{Infinite flag in a channel \label{sec:InfFlag}}

We now consider a simplifed problem for which analytical stability results can be 
obtained. We replace the finite flag with an 
infinite flag, extending upstream and downstream. An example of the infinite flag 
in a channel (thick black line) and image system of flags (gray lines) is shown in
Fig. \ref{fig:SchematicFlagChannel}b. For small deflections, we can linearize
the model equations about the horizontal-flag-in-a-uniform-flow state. The linearized equations
have solutions which are complex exponentials in space and time and 
give a dispersion relation which determines
when a perturbation of a given wavenumber is unstable. This model was used previously
for an infinite flag in unbounded flow \cite{SVZ2005,alben2008ffi}. Now we extend this work to
the case of a channel-bounded infinite flag.

\begin{figure}
  \centerline{\includegraphics[width=10cm]
  {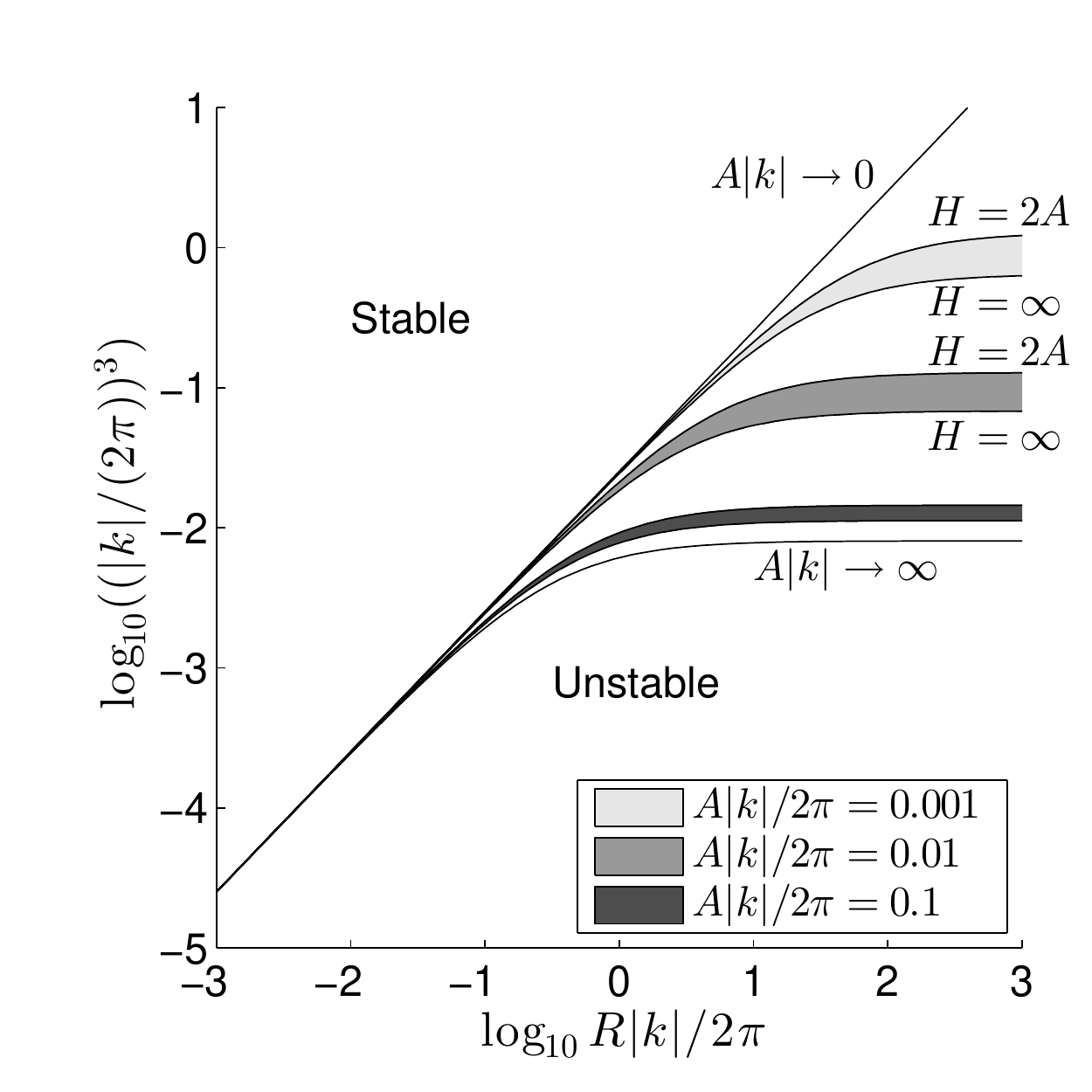}}
  \caption{Plots of stability boundaries for sinusoidal perturbations
of infinite straight flags, in
the space of dimensionless wavenumber and body inertia. Results
are shown for three ratios of near-wall-distance to perturbation 
wavelength ($A|k|/2\pi$). For each value of $A|k|/2\pi$,
a range of stability boundaries is obtained for
$2 \leq H/A < \infty$, shown by a shaded region.}
\label{fig:InfiniteFlagStabilityBoundariesFig}
\end{figure}

We again nondimensionalize velocities by $U$ and fluid densities by $\rho_f$. Since the flag
is infinite, we redefine $L$ (for the infinite flag only) 
to be an intrinsic length based on $B$ and the other parameters in $R_2$:
\begin{align}
L = \left(B/\rho_f U^2 W\right)^{1/3}; R = \frac{\rho_s h}{\rho_f L} \label{InfParams}
\end{align}
\nn Since $L$ is new, we have also renamed $R_1$ with this new $L$ as $R$ in (\ref{InfParams}).
Now there is no free vortex sheet wake, but only a bound vortex sheet on the infinite flag.
We expand equations (\ref{KinematicEqn}), (\ref{Bernoulli}), and (\ref{beam}) about
the horizontal-flag-in-a-uniform-flow state. Then $\zeta(s,t) \approx x + i y(x,t)$,
$\theta(s,t) \approx \partial_x y(x,t)$, $\gamma(s,t) \approx \gamma(x,t)$, etc. 
Equations (\ref{KinematicEqn}), (\ref{Bernoulli}), and (\ref{beam}) reduce to
\begin{align}
R \partial_{tt} y &= - \partial_{xxxx}y - [p], \label{beamInf} \\
\partial_t y + \partial_x y &= \frac{1}{2\pi}\sum_{n = -\infty}^\infty
\int_{-\infty}^{\infty}
\frac{\gamma(x',t)(x-x') \, dx'}{(x-x')^2+ (2nH)^2} 
- \frac{1}{2\pi}\sum_{n = -\infty}^\infty
\int_{-\infty}^{\infty}
\frac{\gamma(x',t)(x-x') \,dx'}{(x-x')^2+(2(nH-A))^2}, \label{KinInf}\\
\partial_t \gamma + \partial_x \gamma &= \partial_x [p].\label{BernInf}
\end{align}
\nn In (\ref{KinInf}), we have expanded the $\coth$ kernels in (\ref{KinematicEqn})
as infinite sums of rational-function kernels. The integrals are somewhat 
easier to evaluate
with these kernels. We now insert complex exponential modes
\begin{align}
y(x,t) = \hat{y}e^{i(kx+\omega t)}\;,\;
\gamma(x,t) = \hat{\gamma}e^{i(kx+\omega t)}\;,\;
[p](x,t) = \hat{[p]}e^{i(kx+\omega t)}
\end{align}
\nn into (\ref{beamInf})--(\ref{BernInf}). 
The integrals in (\ref{KinInf}) are evaluated using the formula
\begin{align}
\frac{1}{2\pi}
\int_{-\infty}^{\infty}
\frac{e^{ikx'}(x-x') \, dx'}{(x-x')^2+ c^2} = 
\frac{-i}{2}\mbox{sgn}(k)e^{-|ck|}e^{ikx}
\end{align}
\nn which can be derived using a complex residue integral. Then
(\ref{KinInf}) becomes
\begin{align}
\hat{y}i(\omega + k) = \frac{-i}{2}\mbox{sgn}(k)\hat{\gamma}
\sum_{n = -\infty}^\infty e^{-|2nHk|} - e^{-|2(nH-A)k|}
\end{align}
\nn in which the sums are geometric series: 
\begin{align}
\sum_{n = -\infty}^\infty e^{-|2nHk|}
&= 1 + 2\frac{e^{-2H|k|}}{1-e^{-2H|k|}} \\
\sum_{n = -\infty}^\infty e^{-|2(nH-A)k|}
&=e^{-2A|k|} + \left(e^{-2A|k|} + e^{2A|k|}\right)
\frac{e^{-2H|k|}}{1-e^{-2H|k|}} \\
\sum_{n = -\infty}^\infty e^{-|2nHk|} - e^{-|2(nH-A)k|}
&= 
\frac{\left(1-e^{-2A|k|}\right)\left(1-e^{-2(H-A)|k|}\right)}
{1-e^{-2H|k|}} \equiv C_{A,H}(k). \\
0 \leq C_{A,H}(k) \leq 1&\;,\;C_{A,H}(k) \to 1 \;\mbox{as}\; A,H \to \infty.
\end{align}
\nn Using (\ref{beamInf})--(\ref{BernInf}) we obtain the dispersion
relation between $\omega$ and $k$:
\begin{align}
\omega^2\left(R |k| + \frac{2}{C_{A,H}(k)}\right) + \omega \frac{4k}{C_{A,H}(k)}
+ \frac{2k^2}{C_{A,H}(k)} - |k|^5 &= 0.
\end{align}
\nn The mode with wavenumber $k$ is unstable (Im$(\omega) < 0$) when
\begin{align}
2R - C_{A,H}(k) R |k|^3 - 2 |k|^2 &> 0 \label{stab}
\end{align}
\nn The stability boundary is a curve in ($R$,$|k|$) space given by
(\ref{stab}) with the inequality replaced by an equality. In order
to compare with the finite-flag
case, we plot the stability boundary in the space of $R|k|/2\pi$ and 
$(|k|/(2\pi))^3$ in figure \ref{fig:InfiniteFlagStabilityBoundariesFig}.

These parameters are used because $|k|^3$ scales with bending rigidity
$B$ in the same way as $R_2$, and $R|k|$ scales with flag and fluid density
in the same way as $R_1$. This can be seen by denoting the dimensional wavenumber by
$\bar{k}$ and dimensional wavelength by $\lambda = 2\pi/|\bar{k}|$, so that
\begin{align}
\frac{R|k|}{2\pi} = \frac{RL|\bar{k}|}{2\pi} = \frac{\rho_s h}{\rho_f \lambda}
\; ; \;
\left(\frac{|k|}{2\pi}\right)^3 = \left(\frac{|\bar{k}|L}{2\pi}\right)^3 = 
\frac{B}{\rho_f U^2 W \lambda^3}
\end{align}
\nn These parameters are equivalent to $R_1$ and $R_2$ when the finite
flag length is replaced with the wavelength of the perturbation of the 
infinite flag.

In figure \ref{fig:InfiniteFlagStabilityBoundariesFig} we plot
stability boundaries in the space of 
$R|k|/2\pi$ and $(|k|/(2\pi))^3$, for different dimensionless
channel spacings, given by the near-wall-spacing
$A|k|/2\pi = A/\lambda$ and the channel width ratio
$H/A$. The lowest line, labeled $A|k| \to \infty$, is
for an unbounded flow, and agrees with previous results
\cite{SVZ2005,AS2008}. Above the line, the flat state is stable,
and below the line it is unstable. Above this line are the stability
boundaries for channel-bounded flows. The shaded regions are
a continuous range of stability boundaries for a given value
of $A|k|/2\pi$ (labeled in the figure legend) and the full range of
possible $H/A$ from 2 to $\infty$. As $A|k|/2\pi$ decreases,
the shaded regions move upward, meaning the flag becomes less
stable, as was found for the finite flag. 
And as $H/A$ decreases from $\infty$ down to 2, 
the stability boundary moves upward through each shaded region.
Both results can be summarized with the statement that 
for ``heavier''  flags ($R|k|/2\pi \gtrsim 1$), the
narrower the channel, the less stable the flag, in agreement
with the finite flag. 
As $R|k|/2\pi$ decreases, all the stability boundaries 
converge. For the finite flag, the stability boundaries in
Fig. \ref{fig:StabBdiesFig}
are quite close at small $R_1$ (the analog of $R|k|/2\pi$), 
but there is a slight
variation that correlates with the channel spacing.

When $A|k| \to \infty$, the stability boundary is similar
to those for the unbounded finite flag in Fig. 
\ref{fig:StabBdiesFig}, agreeing quantitatively at smaller 
inertia (if ($R_1, R_2$) is identified with 
($R|k|/2\pi$,$|k|^3/(2\pi)^3$)) and 
also reaching a plateau at larger inertia, though the
values at which the plateaus occur in $R_2$ and $|k|^3/(2\pi)^3$ 
differ by about one decade. Apparently the presence of the wake
is less important with smaller-wavelength perturbations. This
was found previously for the unbounded flag\cite{AS2008}
and for flapping foils\cite{Alben2008JFM1} .

\section{Conclusion \label{sec:Conc}}

In this work we have investigated some of the basic effects of
channel confinement on the flag flutter instability and large-amplitude
flapping dynamics. For the infinite flag, we found a closed-form
expression for the stability boundaries. In general, greater confinement
increases the size of the instability region, although the effect mainly occurs
with heavier bodies, and disappears for lighter
bodies. For the finite flag, channel confinement has a similar
effect on the stability boundary location. We found that
multiple large-amplitude flapping states are possible but rare
at parameters where periodic dynamics are found.
We quantified the large-amplitude dynamics with three quantities:
the flapping amplitude, the dominant frequency, and the time-averaged
number of extrema in the flag's deflection (analogous to the
wavenumber). We studied the behavior of these quantities as the near-wall
distance decreases, for symmetric and asymmetric channels. Starting
from its unbounded value, the amplitude increases slightly as the
channel walls approach in a symmetric channel, then 
decreases roughly in proportion to the
channel wall spacing. The amplitude may be extremely close to the
near-wall distance ($<$ 0.1 \% in relative magnitude), or much
smaller (20-50\%), when the channel walls are much smaller than the flag
length. 
Meanwhile, the frequency and number of extrema
undergo a sequence of slight increases punctuated by large jumps.
As the wall spacing decreases, the flag amplitude may decrease
greatly with little change in the frequency and number of extrema.
But at certain wall spacing values, the flag jumps to a higher 
flapping mode. The results are similar for an a very 
asymmetric channel, except that the flag flaps with a larger 
amplitude on the less bounded side. 

An important future work is to compare these results with
simulations of the fully viscous flow, to understand how 
the flag-wall interaction is modified 
by the presence of a viscous boundary layer. 

\begin{acknowledgments}
We acknowledge helpful discussions with 
Ari Glezer and Rajat Mittal, who
suggested this problem,
and support from 
Air Force Office of Scientific Research Grant RD547-G2.
\end{acknowledgments}
\bibliography{FlagInChannel}

\begin{thebibliography}{34}%
\makeatletter
\providecommand \@ifxundefined [1]{%
 \@ifx{#1\undefined}
}%
\providecommand \@ifnum [1]{%
 \ifnum #1\expandafter \@firstoftwo
 \else \expandafter \@secondoftwo
 \fi
}%
\providecommand \@ifx [1]{%
 \ifx #1\expandafter \@firstoftwo
 \else \expandafter \@secondoftwo
 \fi
}%
\providecommand \natexlab [1]{#1}%
\providecommand \enquote  [1]{``#1''}%
\providecommand \bibnamefont  [1]{#1}%
\providecommand \bibfnamefont [1]{#1}%
\providecommand \citenamefont [1]{#1}%
\providecommand \href@noop [0]{\@secondoftwo}%
\providecommand \href [0]{\begingroup \@sanitize@url \@href}%
\providecommand \@href[1]{\@@startlink{#1}\@@href}%
\providecommand \@@href[1]{\endgroup#1\@@endlink}%
\providecommand \@sanitize@url [0]{\catcode `\\12\catcode `\$12\catcode
  `\&12\catcode `\#12\catcode `\^12\catcode `\_12\catcode `\%12\relax}%
\providecommand \@@startlink[1]{}%
\providecommand \@@endlink[0]{}%
\providecommand \url  [0]{\begingroup\@sanitize@url \@url }%
\providecommand \@url [1]{\endgroup\@href {#1}{\urlprefix }}%
\providecommand \urlprefix  [0]{URL }%
\providecommand \Eprint [0]{\href }%
\providecommand \doibase [0]{http://dx.doi.org/}%
\providecommand \selectlanguage [0]{\@gobble}%
\providecommand \bibinfo  [0]{\@secondoftwo}%
\providecommand \bibfield  [0]{\@secondoftwo}%
\providecommand \translation [1]{[#1]}%
\providecommand \BibitemOpen [0]{}%
\providecommand \bibitemStop [0]{}%
\providecommand \bibitemNoStop [0]{.\EOS\space}%
\providecommand \EOS [0]{\spacefactor3000\relax}%
\providecommand \BibitemShut  [1]{\csname bibitem#1\endcsname}%
\let\auto@bib@innerbib\@empty
\bibitem [{\citenamefont {Kornecki}, \citenamefont {Dowell},\ and\
  \citenamefont {O'Brien}(1976)}]{kornecki1976ait}%
  \BibitemOpen
  \bibfield  {author} {\bibinfo {author} {\bibfnamefont {A.}~\bibnamefont
  {Kornecki}}, \bibinfo {author} {\bibfnamefont {E.}~\bibnamefont {Dowell}}, \
  and\ \bibinfo {author} {\bibfnamefont {J.}~\bibnamefont {O'Brien}},\
  }\bibfield  {title} {\enquote {\bibinfo {title} {{On the aeroelastic
  instability of two-dimensional panels in uniform incompressible flow}},}\
  }\href@noop {} {\bibfield  {journal} {\bibinfo  {journal} {J. Sound
  Vibration}\ }\textbf {\bibinfo {volume} {47}},\ \bibinfo {pages} {163--178}
  (\bibinfo {year} {1976})}\BibitemShut {NoStop}%
\bibitem [{\citenamefont {Dowell}(1980)}]{Dowell1980}%
  \BibitemOpen
  \bibfield  {author} {\bibinfo {author} {\bibfnamefont {E.}~\bibnamefont
  {Dowell}},\ }\bibfield  {title} {\enquote {\bibinfo {title} {Nonlinear
  aeroelasticity},}\ }in\ \href@noop {} {\emph {\bibinfo {booktitle} {New
  approaches to nonlinear problems in dynamics}}}\ (\bibinfo  {publisher} {SIAM
  Publications},\ \bibinfo {year} {1980})\ pp.\ \bibinfo {pages}
  {147--172}\BibitemShut {NoStop}%
\bibitem [{\citenamefont {Huang}(1995)}]{Huang_JFluidsStruct_1995}%
  \BibitemOpen
  \bibfield  {author} {\bibinfo {author} {\bibfnamefont {L.}~\bibnamefont
  {Huang}},\ }\bibfield  {title} {\enquote {\bibinfo {title} {Flutter of
  cantilevered plates in axial flow},}\ }\href@noop {} {\bibfield  {journal}
  {\bibinfo  {journal} {J. Fluids Struct.}\ }\textbf {\bibinfo {volume} {9}},\
  \bibinfo {pages} {127--147} (\bibinfo {year} {1995})}\BibitemShut {NoStop}%
\bibitem [{\citenamefont {Zhang}\ \emph {et~al.}(2000)\citenamefont {Zhang},
  \citenamefont {Childress}, \citenamefont {Libchaber},\ and\ \citenamefont
  {Shelley}}]{ZCLS2000}%
  \BibitemOpen
  \bibfield  {author} {\bibinfo {author} {\bibfnamefont {J.}~\bibnamefont
  {Zhang}}, \bibinfo {author} {\bibfnamefont {S.}~\bibnamefont {Childress}},
  \bibinfo {author} {\bibfnamefont {A.}~\bibnamefont {Libchaber}}, \ and\
  \bibinfo {author} {\bibfnamefont {M.}~\bibnamefont {Shelley}},\ }\bibfield
  {title} {\enquote {\bibinfo {title} {Flexible filaments in a flowing soap
  film as a model for flags in a two dimensional wind},}\ }\href@noop {}
  {\bibfield  {journal} {\bibinfo  {journal} {Nature}\ }\textbf {\bibinfo
  {volume} {408}},\ \bibinfo {pages} {835--839} (\bibinfo {year}
  {2000})}\BibitemShut {NoStop}%
\bibitem [{\citenamefont {Fitt}\ and\ \citenamefont {Pope}(2001)}]{FP2001}%
  \BibitemOpen
  \bibfield  {author} {\bibinfo {author} {\bibfnamefont {A.}~\bibnamefont
  {Fitt}}\ and\ \bibinfo {author} {\bibfnamefont {M.}~\bibnamefont {Pope}},\
  }\bibfield  {title} {\enquote {\bibinfo {title} {The unsteady motion of
  two-dimensional flags with bending stiffness},}\ }\href@noop {} {\bibfield
  {journal} {\bibinfo  {journal} {J. Eng. Math.}\ }\textbf {\bibinfo {volume}
  {40}},\ \bibinfo {pages} {227--248} (\bibinfo {year} {2001})}\BibitemShut
  {NoStop}%
\bibitem [{\citenamefont {Watanabe}\ \emph {et~al.}(2002)\citenamefont
  {Watanabe}, \citenamefont {Suzuki}, \citenamefont {Sugihara},\ and\
  \citenamefont {Sueoka}}]{watanabe2002esp}%
  \BibitemOpen
  \bibfield  {author} {\bibinfo {author} {\bibfnamefont {Y.}~\bibnamefont
  {Watanabe}}, \bibinfo {author} {\bibfnamefont {S.}~\bibnamefont {Suzuki}},
  \bibinfo {author} {\bibfnamefont {M.}~\bibnamefont {Sugihara}}, \ and\
  \bibinfo {author} {\bibfnamefont {Y.}~\bibnamefont {Sueoka}},\ }\bibfield
  {title} {\enquote {\bibinfo {title} {{An experimental study of paper
  flutter}},}\ }\href@noop {} {\bibfield  {journal} {\bibinfo  {journal} {J.
  Fluids Struct.}\ }\textbf {\bibinfo {volume} {16}},\ \bibinfo {pages}
  {529--542} (\bibinfo {year} {2002})}\BibitemShut {NoStop}%
\bibitem [{\citenamefont {Zhu}\ and\ \citenamefont {Peskin}(2002)}]{ZP2002}%
  \BibitemOpen
  \bibfield  {author} {\bibinfo {author} {\bibfnamefont {L.}~\bibnamefont
  {Zhu}}\ and\ \bibinfo {author} {\bibfnamefont {C.}~\bibnamefont {Peskin}},\
  }\bibfield  {title} {\enquote {\bibinfo {title} {Simulation of a flapping
  flexible filament in a flowing soap film by the immersed boundary method},}\
  }\href@noop {} {\bibfield  {journal} {\bibinfo  {journal} {J. Comput. Phys.}\
  }\textbf {\bibinfo {volume} {{179}}},\ \bibinfo {pages} {452--468} (\bibinfo
  {year} {2002})}\BibitemShut {NoStop}%
\bibitem [{\citenamefont {Tang}, \citenamefont {Yamamoto},\ and\ \citenamefont
  {Dowell}(2003)}]{TYD_JFluidsStruct_2003}%
  \BibitemOpen
  \bibfield  {author} {\bibinfo {author} {\bibfnamefont {D.}~\bibnamefont
  {Tang}}, \bibinfo {author} {\bibfnamefont {H.}~\bibnamefont {Yamamoto}}, \
  and\ \bibinfo {author} {\bibfnamefont {E.}~\bibnamefont {Dowell}},\
  }\bibfield  {title} {\enquote {\bibinfo {title} {Flutter and limit cycle
  oscillations of two-dimensional panels in three-dimensional axial flow},}\
  }\href@noop {} {\bibfield  {journal} {\bibinfo  {journal} {Journal of Fluids
  and Structures}\ }\textbf {\bibinfo {volume} {17}},\ \bibinfo {pages}
  {225--242} (\bibinfo {year} {2003})}\BibitemShut {NoStop}%
\bibitem [{\citenamefont {Shelley}, \citenamefont {Vandenberghe},\ and\
  \citenamefont {Zhang}(2005)}]{SVZ2005}%
  \BibitemOpen
  \bibfield  {author} {\bibinfo {author} {\bibfnamefont {M.}~\bibnamefont
  {Shelley}}, \bibinfo {author} {\bibfnamefont {N.}~\bibnamefont
  {Vandenberghe}}, \ and\ \bibinfo {author} {\bibfnamefont {J.}~\bibnamefont
  {Zhang}},\ }\bibfield  {title} {\enquote {\bibinfo {title} {Heavy flags
  undergo spontaneous oscillations in flowing water},}\ }\href@noop {}
  {\bibfield  {journal} {\bibinfo  {journal} {Phys. Rev. Lett}\ }\textbf
  {\bibinfo {volume} {94}},\ \bibinfo {pages} {094302} (\bibinfo {year}
  {2005})}\BibitemShut {NoStop}%
\bibitem [{\citenamefont {Argentina}\ and\ \citenamefont
  {Mahadevan}(2005)}]{AM2005}%
  \BibitemOpen
  \bibfield  {author} {\bibinfo {author} {\bibfnamefont {M.}~\bibnamefont
  {Argentina}}\ and\ \bibinfo {author} {\bibfnamefont {L.}~\bibnamefont
  {Mahadevan}},\ }\bibfield  {title} {\enquote {\bibinfo {title}
  {{Fluid-flow-induced flutter of a flag}},}\ }\href@noop {} {\bibfield
  {journal} {\bibinfo  {journal} {Proc. Natl. Acad. Sci. (USA)}\ }\textbf
  {\bibinfo {volume} {102}},\ \bibinfo {pages} {1829--1834} (\bibinfo {year}
  {2005})}\BibitemShut {NoStop}%
\bibitem [{\citenamefont {Eloy}, \citenamefont {Souilliez},\ and\ \citenamefont
  {Schouveiler}(2007)}]{ESS2007}%
  \BibitemOpen
  \bibfield  {author} {\bibinfo {author} {\bibfnamefont {C.}~\bibnamefont
  {Eloy}}, \bibinfo {author} {\bibfnamefont {C.}~\bibnamefont {Souilliez}}, \
  and\ \bibinfo {author} {\bibfnamefont {L.}~\bibnamefont {Schouveiler}},\
  }\bibfield  {title} {\enquote {\bibinfo {title} {Flutter of a rectangular
  plate},}\ }\href@noop {} {\bibfield  {journal} {\bibinfo  {journal} {J.
  Fluid. Struct.}\ }\textbf {\bibinfo {volume} {23}},\ \bibinfo {pages}
  {904--919} (\bibinfo {year} {2007})}\BibitemShut {NoStop}%
\bibitem [{\citenamefont {Connell}\ and\ \citenamefont
  {Yue}(2007)}]{connell2007fdf}%
  \BibitemOpen
  \bibfield  {author} {\bibinfo {author} {\bibfnamefont {B.}~\bibnamefont
  {Connell}}\ and\ \bibinfo {author} {\bibfnamefont {D.}~\bibnamefont {Yue}},\
  }\bibfield  {title} {\enquote {\bibinfo {title} {Flapping dynamics of a flag
  in a uniform stream},}\ }\href@noop {} {\bibfield  {journal} {\bibinfo
  {journal} {J. Fluid Mech.}\ }\textbf {\bibinfo {volume} {581}},\ \bibinfo
  {pages} {33--67} (\bibinfo {year} {2007})}\BibitemShut {NoStop}%
\bibitem [{\citenamefont {Eloy}\ \emph {et~al.}(2008)\citenamefont {Eloy},
  \citenamefont {Lagrange}, \citenamefont {Souilliez},\ and\ \citenamefont
  {Schouveiler}}]{eloy2008aic}%
  \BibitemOpen
  \bibfield  {author} {\bibinfo {author} {\bibfnamefont {C.}~\bibnamefont
  {Eloy}}, \bibinfo {author} {\bibfnamefont {R.}~\bibnamefont {Lagrange}},
  \bibinfo {author} {\bibfnamefont {C.}~\bibnamefont {Souilliez}}, \ and\
  \bibinfo {author} {\bibfnamefont {L.}~\bibnamefont {Schouveiler}},\
  }\bibfield  {title} {\enquote {\bibinfo {title} {{Aeroelastic instability of
  cantilevered flexible plates in uniform flow}},}\ }\href@noop {} {\bibfield
  {journal} {\bibinfo  {journal} {Journal of Fluid Mechanics}\ }\textbf
  {\bibinfo {volume} {611}},\ \bibinfo {pages} {97--106} (\bibinfo {year}
  {2008})}\BibitemShut {NoStop}%
\bibitem [{\citenamefont {Alben}(2008{\natexlab{a}})}]{alben2008ffi}%
  \BibitemOpen
  \bibfield  {author} {\bibinfo {author} {\bibfnamefont {S.}~\bibnamefont
  {Alben}},\ }\bibfield  {title} {\enquote {\bibinfo {title} {{The
  flapping-flag instability as a nonlinear eigenvalue problem}},}\ }\href@noop
  {} {\bibfield  {journal} {\bibinfo  {journal} {Physics of Fluids}\ }\textbf
  {\bibinfo {volume} {20}},\ \bibinfo {pages} {104106} (\bibinfo {year}
  {2008}{\natexlab{a}})}\BibitemShut {NoStop}%
\bibitem [{\citenamefont {Michelin}, \citenamefont {Smith},\ and\ \citenamefont
  {Glover}(2008)}]{Michelin2008}%
  \BibitemOpen
  \bibfield  {author} {\bibinfo {author} {\bibfnamefont {S.}~\bibnamefont
  {Michelin}}, \bibinfo {author} {\bibfnamefont {S.~G.~L.}\ \bibnamefont
  {Smith}}, \ and\ \bibinfo {author} {\bibfnamefont {B.~J.}\ \bibnamefont
  {Glover}},\ }\bibfield  {title} {\enquote {\bibinfo {title} {{Vortex shedding
  model of a flapping flag}},}\ }\href@noop {} {\bibfield  {journal} {\bibinfo
  {journal} {{Journal of Fluid Mechanics}}\ }\textbf {\bibinfo {volume}
  {617}},\ \bibinfo {pages} {1--10} (\bibinfo {year} {2008})}\BibitemShut
  {NoStop}%
\bibitem [{\citenamefont {Manela}\ and\ \citenamefont
  {Howe}(2009)}]{manela2009}%
  \BibitemOpen
  \bibfield  {author} {\bibinfo {author} {\bibfnamefont {A.}~\bibnamefont
  {Manela}}\ and\ \bibinfo {author} {\bibfnamefont {M.~S.}\ \bibnamefont
  {Howe}},\ }\bibfield  {title} {\enquote {\bibinfo {title} {{The forced motion
  of a flag}},}\ }\href@noop {} {\bibfield  {journal} {\bibinfo  {journal}
  {{Journal of Fluid Mechanics}}\ }\textbf {\bibinfo {volume} {{635}}},\
  \bibinfo {pages} {439--454} (\bibinfo {year} {{2009}})}\BibitemShut {NoStop}%
\bibitem [{\citenamefont {Theodorsen}(1935)}]{Theodorsen1935}%
  \BibitemOpen
  \bibfield  {author} {\bibinfo {author} {\bibfnamefont {T.}~\bibnamefont
  {Theodorsen}},\ }\href@noop {} {\enquote {\bibinfo {title} {General theory of
  aerodynamic instability and the mechanism of flutter},}\ }\bibinfo {type}
  {Tech. Rep.}\ \bibinfo {number} {496}\ (\bibinfo  {institution} {NACA},\
  \bibinfo {year} {1935})\BibitemShut {NoStop}%
\bibitem [{\citenamefont {Fung}(1955)}]{Fung1955}%
  \BibitemOpen
  \bibfield  {author} {\bibinfo {author} {\bibfnamefont {Y.}~\bibnamefont
  {Fung}},\ }\href@noop {} {\emph {\bibinfo {title} {An Introduction to the
  Theory of Aeroelasticity}}}\ (\bibinfo  {publisher} {Dover Publications},\
  \bibinfo {address} {New York},\ \bibinfo {year} {1955})\BibitemShut {NoStop}%
\bibitem [{\citenamefont {Bisplinghoff}\ and\ \citenamefont
  {Ashley}(2002)}]{bisplinghoff2002pa}%
  \BibitemOpen
  \bibfield  {author} {\bibinfo {author} {\bibfnamefont {R.}~\bibnamefont
  {Bisplinghoff}}\ and\ \bibinfo {author} {\bibfnamefont {H.}~\bibnamefont
  {Ashley}},\ }\href@noop {} {\emph {\bibinfo {title} {{Principles of
  Aeroelasticity}}}}\ (\bibinfo  {publisher} {Dover Pub.},\ \bibinfo {address}
  {New York},\ \bibinfo {year} {2002})\BibitemShut {NoStop}%
\bibitem [{\citenamefont {Shelley}\ and\ \citenamefont
  {Zhang}(2011)}]{shelley2011flapping}%
  \BibitemOpen
  \bibfield  {author} {\bibinfo {author} {\bibfnamefont {M.}~\bibnamefont
  {Shelley}}\ and\ \bibinfo {author} {\bibfnamefont {J.}~\bibnamefont
  {Zhang}},\ }\bibfield  {title} {\enquote {\bibinfo {title} {{Flapping and
  Bending Bodies Interacting with Fluid Flows}},}\ }\href@noop {} {\bibfield
  {journal} {\bibinfo  {journal} {Annual Review of Fluid Mechanics}\ }\textbf
  {\bibinfo {volume} {43}},\ \bibinfo {pages} {449--465} (\bibinfo {year}
  {2011})}\BibitemShut {NoStop}%
\bibitem [{\citenamefont {Alben}\ and\ \citenamefont {Shelley}(2008)}]{AS2008}%
  \BibitemOpen
  \bibfield  {author} {\bibinfo {author} {\bibfnamefont {S.}~\bibnamefont
  {Alben}}\ and\ \bibinfo {author} {\bibfnamefont {M.}~\bibnamefont
  {Shelley}},\ }\bibfield  {title} {\enquote {\bibinfo {title} {Flapping states
  of a flag in an inviscid fluid: Bistability and the transition to chaos},}\
  }\href@noop {} {\bibfield  {journal} {\bibinfo  {journal} {Phys. Rev. Lett.}\
  }\textbf {\bibinfo {volume} {100}},\ \bibinfo {pages} {074301} (\bibinfo
  {year} {2008})}\BibitemShut {NoStop}%
\bibitem [{\citenamefont {Chen}\ \emph {et~al.}(2014)\citenamefont {Chen},
  \citenamefont {Jia}, \citenamefont {Wu}, \citenamefont {Yin},\ and\
  \citenamefont {Ma}}]{chen2014bifurcation}%
  \BibitemOpen
  \bibfield  {author} {\bibinfo {author} {\bibfnamefont {M.}~\bibnamefont
  {Chen}}, \bibinfo {author} {\bibfnamefont {L.-B.}\ \bibnamefont {Jia}},
  \bibinfo {author} {\bibfnamefont {Y.-F.}\ \bibnamefont {Wu}}, \bibinfo
  {author} {\bibfnamefont {X.-Z.}\ \bibnamefont {Yin}}, \ and\ \bibinfo
  {author} {\bibfnamefont {Y.-B.}\ \bibnamefont {Ma}},\ }\bibfield  {title}
  {\enquote {\bibinfo {title} {Bifurcation and chaos of a flag in an inviscid
  flow},}\ }\href@noop {} {\bibfield  {journal} {\bibinfo  {journal} {Journal
  of Fluids and Structures}\ }\textbf {\bibinfo {volume} {45}},\ \bibinfo
  {pages} {124--137} (\bibinfo {year} {2014})}\BibitemShut {NoStop}%
\bibitem [{\citenamefont {Doare}, \citenamefont {Mano},\ and\ \citenamefont
  {Ludena}(2011)}]{Doare2011}%
  \BibitemOpen
  \bibfield  {author} {\bibinfo {author} {\bibfnamefont {O.}~\bibnamefont
  {Doare}}, \bibinfo {author} {\bibfnamefont {D.}~\bibnamefont {Mano}}, \ and\
  \bibinfo {author} {\bibfnamefont {J.~C.~B.}\ \bibnamefont {Ludena}},\
  }\bibfield  {title} {\enquote {\bibinfo {title} {{Effect of spanwise
  confinement on flag flutter: Experimental measurements}},}\ }\href@noop {}
  {\bibfield  {journal} {\bibinfo  {journal} {{Physics of Fluids}}\ }\textbf
  {\bibinfo {volume} {{23}}} (\bibinfo {year} {{2011}})}\BibitemShut {NoStop}%
\bibitem [{\citenamefont {Doare}, \citenamefont {Sauzade},\ and\ \citenamefont
  {Eloy}(2011)}]{Doare2011a}%
  \BibitemOpen
  \bibfield  {author} {\bibinfo {author} {\bibfnamefont {O.}~\bibnamefont
  {Doare}}, \bibinfo {author} {\bibfnamefont {M.}~\bibnamefont {Sauzade}}, \
  and\ \bibinfo {author} {\bibfnamefont {C.}~\bibnamefont {Eloy}},\ }\bibfield
  {title} {\enquote {\bibinfo {title} {{Flutter of an elastic plate in a
  channel flow: Confinement and finite-size effects}},}\ }\href@noop {}
  {\bibfield  {journal} {\bibinfo  {journal} {{Journal of Fluids and
  Structures}}\ }\textbf {\bibinfo {volume} {27}},\ \bibinfo {pages} {76--88}
  (\bibinfo {year} {2011})}\BibitemShut {NoStop}%
\bibitem [{\citenamefont {Guo}\ and\ \citenamefont
  {Paidoussis}(2000)}]{Guo2000}%
  \BibitemOpen
  \bibfield  {author} {\bibinfo {author} {\bibfnamefont {C.}~\bibnamefont
  {Guo}}\ and\ \bibinfo {author} {\bibfnamefont {M.}~\bibnamefont
  {Paidoussis}},\ }\bibfield  {title} {\enquote {\bibinfo {title} {{Stability
  of rectangular plates with free side-edges in two-dimensional inviscid
  channel flow}},}\ }\href@noop {} {\bibfield  {journal} {\bibinfo  {journal}
  {{Journal of Applied Mechanics-Transactions of the ASME}}\ }\textbf {\bibinfo
  {volume} {{67}}},\ \bibinfo {pages} {171--176} (\bibinfo {year}
  {{2000}})}\BibitemShut {NoStop}%
\bibitem [{\citenamefont {Jaiman}, \citenamefont {Parmar},\ and\ \citenamefont
  {Gurugubelli}(2014)}]{Jaiman2014}%
  \BibitemOpen
  \bibfield  {author} {\bibinfo {author} {\bibfnamefont {R.~K.}\ \bibnamefont
  {Jaiman}}, \bibinfo {author} {\bibfnamefont {M.~K.}\ \bibnamefont {Parmar}},
  \ and\ \bibinfo {author} {\bibfnamefont {P.~S.}\ \bibnamefont
  {Gurugubelli}},\ }\bibfield  {title} {\enquote {\bibinfo {title} {{Added Mass
  and Aeroelastic Stability of a Flexible Plate Interacting With Mean Flow in a
  Confined Channel}},}\ }\href@noop {} {\bibfield  {journal} {\bibinfo
  {journal} {{Journal of Applied Mechanics-Transactions of the ASME}}\ }\textbf
  {\bibinfo {volume} {{81}}} (\bibinfo {year} {{2014}})}\BibitemShut {NoStop}%
\bibitem [{\citenamefont {Alben}(2009)}]{AlbenJCP2009}%
  \BibitemOpen
  \bibfield  {author} {\bibinfo {author} {\bibfnamefont {S.}~\bibnamefont
  {Alben}},\ }\bibfield  {title} {\enquote {\bibinfo {title} {{Simulating the
  dynamics of flexible bodies and vortex sheets}},}\ }\href@noop {} {\bibfield
  {journal} {\bibinfo  {journal} {{J. Comp. Phys.}}\ }\textbf {\bibinfo
  {volume} {{228}}},\ \bibinfo {pages} {2587--2603} (\bibinfo {year}
  {{2009}})}\BibitemShut {NoStop}%
\bibitem [{\citenamefont {Saffman}(1992)}]{Saffman1992}%
  \BibitemOpen
  \bibfield  {author} {\bibinfo {author} {\bibfnamefont {P.}~\bibnamefont
  {Saffman}},\ }\href@noop {} {\emph {\bibinfo {title} {Vortex Dynamics}}}\
  (\bibinfo  {publisher} {Cambridge Univ. Press},\ \bibinfo {address}
  {Cambridge},\ \bibinfo {year} {1992})\BibitemShut {NoStop}%
\bibitem [{\citenamefont {Krasny}(1986)}]{Krasny1986}%
  \BibitemOpen
  \bibfield  {author} {\bibinfo {author} {\bibfnamefont {R.}~\bibnamefont
  {Krasny}},\ }\bibfield  {title} {\enquote {\bibinfo {title}
  {Desingularization of periodic vortex sheet roll-up},}\ }\href@noop {}
  {\bibfield  {journal} {\bibinfo  {journal} {J. Comp. Phys.}\ }\textbf
  {\bibinfo {volume} {65}},\ \bibinfo {pages} {292--313} (\bibinfo {year}
  {1986})}\BibitemShut {NoStop}%
\bibitem [{\citenamefont {Alben}(2010)}]{alben2010regularizing}%
  \BibitemOpen
  \bibfield  {author} {\bibinfo {author} {\bibfnamefont {S.}~\bibnamefont
  {Alben}},\ }\bibfield  {title} {\enquote {\bibinfo {title} {{Regularizing a
  vortex sheet near a separation point}},}\ }\href@noop {} {\bibfield
  {journal} {\bibinfo  {journal} {Journal of Computational Physics}\ }\textbf
  {\bibinfo {volume} {229}},\ \bibinfo {pages} {5280--5298} (\bibinfo {year}
  {2010})}\BibitemShut {NoStop}%
\bibitem [{\citenamefont {Hou}, \citenamefont {Lowengrub},\ and\ \citenamefont
  {Shelley}(2001)}]{HLS_JComputPhys_2001}%
  \BibitemOpen
  \bibfield  {author} {\bibinfo {author} {\bibfnamefont {T.}~\bibnamefont
  {Hou}}, \bibinfo {author} {\bibfnamefont {J.}~\bibnamefont {Lowengrub}}, \
  and\ \bibinfo {author} {\bibfnamefont {M.}~\bibnamefont {Shelley}},\
  }\bibfield  {title} {\enquote {\bibinfo {title} {Boundary integrals methods
  for multicomponent fluids and multiphase materials},}\ }\href@noop {}
  {\bibfield  {journal} {\bibinfo  {journal} {J. Comput. Phys.}\ }\textbf
  {\bibinfo {volume} {169}},\ \bibinfo {pages} {302--362} (\bibinfo {year}
  {2001})}\BibitemShut {NoStop}%
\bibitem [{\citenamefont {Alben}(2012)}]{alben2012attraction}%
  \BibitemOpen
  \bibfield  {author} {\bibinfo {author} {\bibfnamefont {S.}~\bibnamefont
  {Alben}},\ }\bibfield  {title} {\enquote {\bibinfo {title} {{The attraction
  between a flexible filament and a point vortex}},}\ }\href@noop {} {\bibfield
   {journal} {\bibinfo  {journal} {Journal of Fluid Mechanics}\ }\textbf
  {\bibinfo {volume} {697}},\ \bibinfo {pages} {481--503} (\bibinfo {year}
  {2012})}\BibitemShut {NoStop}%
\bibitem [{\citenamefont {Welch}(1967)}]{welch1967use}%
  \BibitemOpen
  \bibfield  {author} {\bibinfo {author} {\bibfnamefont {P.~D.}\ \bibnamefont
  {Welch}},\ }\bibfield  {title} {\enquote {\bibinfo {title} {The use of fast
  fourier transform for the estimation of power spectra: a method based on time
  averaging over short, modified periodograms},}\ }\href@noop {} {\bibfield
  {journal} {\bibinfo  {journal} {IEEE Transactions on Audio and
  Electroacoustics}\ }\textbf {\bibinfo {volume} {15}},\ \bibinfo {pages}
  {70--73} (\bibinfo {year} {1967})}\BibitemShut {NoStop}%
\bibitem [{\citenamefont {Alben}(2008{\natexlab{b}})}]{Alben2008JFM1}%
  \BibitemOpen
  \bibfield  {author} {\bibinfo {author} {\bibfnamefont {S.}~\bibnamefont
  {Alben}},\ }\bibfield  {title} {\enquote {\bibinfo {title} {{Optimal
  flexibility of a flapping appendage at high Reynolds number}},}\ }\href@noop
  {} {\bibfield  {journal} {\bibinfo  {journal} {J. Fluid Mech.}\ }\textbf
  {\bibinfo {volume} {614}},\ \bibinfo {pages} {355--380} (\bibinfo {year}
  {2008}{\natexlab{b}})}\BibitemShut {NoStop}%
\end{thebibliography}%

\end{document}